\documentclass[conference]{IEEEtran}
\IEEEoverridecommandlockouts
\usepackage{cite}
\usepackage{amsmath,amssymb,amsfonts}
\usepackage{algorithmic}
\usepackage{graphicx}
\usepackage{textcomp}
\usepackage{minted}
\usepackage{xcolor}
\usepackage{pgfplots}
\usepackage{dirtytalk}
\usepackage[inline]{enumitem}
\usepackage{hyperref}
\usepackage{cleveref}
\def\BibTeX{{\rm B\kern-.05em{\sc i\kern-.025em b}\kern-.08em
T\kern-.1667em\lower.7ex\hbox{E}\kern-.125emX}}
\usepackage[table,xcdraw]{xcolor} 

\usepackage{tikz}
\usetikzlibrary{shapes.geometric, shapes.misc, arrows.meta, positioning, calc}

\usepackage{todonotes}

\newcommand{\defectCategory}[1]{{\small\textsf{#1}}}
\newcommand{\keyTerm}[1]{\emph{#1}}
\usepackage{url}

\begin{document}

\title{Benchmarking LLM-Based Static Analysis for Secure Smart Contract Development: Reliability, Limitations, and Potential Hybrid Solutions\thanks{This work is an extended version of the paper accepted for publication at IEEE COMPSAC 2026.}}

\author{\IEEEauthorblockN{1\textsuperscript{st} Susan Ștefan-Claudiu}
\IEEEauthorblockA{\textit{Department of Computer Science} \\
\textit{Alexandru Ioan Cuza University of Iași}\\
Iași, Romania \\
\href{mailto:claudiu\_susan@yahoo.com}{claudiu\_susan@yahoo.com}}
\and
\IEEEauthorblockN{2\textsuperscript{nd} Arusoaie Andrei}
\IEEEauthorblockA{\textit{Department of Computer Science} \\
\textit{Alexandru Ioan Cuza University of Iași}\\
Iași, Romania \\
\href{mailto:andrei.arusoaie@uaic.ro}{andrei.arusoaie@uaic.ro}}
\and
\IEEEauthorblockN{3\textsuperscript{rd} Lucanu Dorel}
\IEEEauthorblockA{\textit{Department of Computer Science} \\
\textit{Alexandru Ioan Cuza University of Iași}\\
Iași, Romania \\
\href{mailto:dorel.lucanu@uaic.ro}{dorel.lucanu@uaic.ro}}
}

\maketitle

\begin{abstract}

The irreversible nature of blockchain transactions makes the identification of smart contract vulnerabilities an essential requirement for secure system development. While Large Language Models (LLMs) are increasingly integrated into developer workflows, their reliability as autonomous security auditors remains unproven.
We assess whether current generative models are a viable replacement for, or only a complement to, traditional static-analysis tools.

Our findings indicate that LLM efficacy is undermined by both inherent lexical bias and a lack of rigorous validation of external data inputs. This reliance on non-semantic heuristics, such as identifier naming, leads to a high frequency of false positives. Furthermore, prompting techniques reveal a trade-off between precision and recall. 
These results were derived using our custom automated framework, which achieves 92\% accuracy 
in classifying model outputs.
\end{abstract}


\section{Introduction}
The implementation of smart contracts on blockchain platforms has introduced a \say{code is law} paradigm, where software vulnerabilities in immutable code lead to financial losses. 
Traditional static analysis tools such as Slither~\cite{slither} and Mythril~\cite{mythril} have long served as the first line of defense; however, they are often constrained by rigid rule sets when faced with complex, non-standard code logic. On the other hand, LLMs offer a promising alternative for automated auditing due to their vast training data and ability to generalize to new smart contract implementations.

Integrating LLMs into the blockchain security lifecycle introduces a theoretical paradox. While these models may be able to identify potential exploits, it remains unclear whether they can mimic genuine semantic reasoning or rely solely on probabilistic pattern matching—a phenomenon known as \keyTerm{lexical bias}. Recent findings from Apple~\cite{AppleReasoningModels}, show that LLM performance degrades notably when encountering superficial variations in problem descriptions, suggesting an overreliance on lexical cues rather than robust logical derivation.

In the context of developing secure smart contracts, a detector that relies on variable names rather than execution logic is not reliable. This reliance (see, e.g.,~\cite{StochasticParrot}) makes an LLM a \say{stochastic auditor}: it can be effective when the code matches its training distribution, but it may fail on novel or sanitized logic that lacks descriptive identifiers. 
This risk is amplified with the shift towards Agentic AI, where LLMs are expected to function as proactive, autonomous agents for decision-making. To investigate the potential for LLMs to augment existing security issues, this study addresses the following research questions: 

\begin{description}[nosep]
    \item [RQ1: Lexical Bias] To what extent do LLMs rely on non-semantic code heuristics, such as descriptive identifier names and comments, rather than semantic execution logic for identifying smart contract vulnerabilities?
    \item[RQ2: Reasoning versus Pattern Matching] Does Chain-of-Thought (CoT) prompting enhance the logical reasoning capabilities of LLMs in vulnerability detection?
    \item [RQ3: Hybrid Integration Performance] How does grounding an LLM with security reports from traditional static analysis tools impact the model's detection range?
    \item [RQ4: Hybrid Integration Bias] Does grounding an LLM with security reports from traditional static analysis tools introduce a measurable bias toward the tool's existing findings?
    \item [RQ5: Architectural Efficiency] What are the performance and reliability trade-offs when deploying LLMs locally using 4-bit quantization compared to full-precision API-based models for security auditing? Local deployment matters because some auditing clients must keep proprietary contracts on premises and therefore cannot send them to third-party APIs, a common situation in enterprise treasuries, regulated DeFi projects, and pre-disclosure audits.
\end{description}

We conduct an exploratory empirical investigation to extract patterns and anti-patterns that can inform hybrid LLM-assisted smart contract auditing solutions.

Smart contracts are an ideal case study for vulnerability detection using static analysis techniques because they are \keyTerm{specialized} programs executed in a distributed environment (the blockchain), where code vulnerabilities can result in significant financial losses. Moreover, once deployed, such programs are public and cannot be modified, making the a priori discovery of vulnerabilities highly desirable. 
Existing vulnerable smart contract benchmarks often contain suggestive names (for variables, functions, even contracts) that could hint to LLMs about what they should look for. Measuring LLM detection performance may be of limited value, since syntactic clues can inflate the results. This motivates our paired-variant protocol (see Approach below).

Answering our research questions requires addressing several challenges: finding \keyTerm{a relevant benchmark for using LLMs in static analysis} -- comparable to the existing ones that were designed for traditional static tools; leveraging LLMs to provide \keyTerm{consistent structured output} -- even when explicitly requested to adhere to a strict response format, structural inconsistencies can still be encountered, or receive answers with a mismatching terminology; dealing with \keyTerm{ambiguity in evaluation criteria} -- what counts as a correct detection can be unclear; tackling \keyTerm{prompt sensitivity and instability} -- LLM behavior varies dramatically, depending on prompt phrasing; acknowledging \keyTerm{scale and cost constraints} -- evaluating many LLMs is resource-intensive.

\paragraph*{Approach} 
To address these challenges, we chose an existing collection of smart contract vulnerabilities with a distinctive characteristic: each test item in this collection consists of a pair of smart contracts. Each pair includes a \keyTerm{positive} variant, which contains a vulnerability, and a \keyTerm{negative} variant, which is very similar to the positive variant but does not include that specific vulnerability. This feature allows us to assess the robustness of LLMs for vulnerability detection, as we can verify whether the detector reports the error in the positive variant and does not report an error in the negative variant. In other words, this collection enables us to evaluate whether, \keyTerm{under very similar conditions}, an AI model can precisely identify vulnerabilities. This makes the collection well-suited for studying the limitations of LLM-based detection.

Our evaluation of the LLMs consists of prompting the models with different techniques, both with and without information from external tools, for each positive/negative variant. We built a reusable infrastructure that allows the experiments to be run consistently and facilitates the integration of additional models.
Since we already have positive/negative variants for each smart contract in the collection, we assign labels to each output to indicate whether it is a \keyTerm{true positive} (TP), \keyTerm{true negative} (TN), \keyTerm{false positive} (FP), or \keyTerm{false negative} (FN). We performed this labeling manually for several tests, but we also developed a method to \keyTerm{automatically} assign these labels with \keyTerm{high} accuracy. 

\paragraph*{Contributions} 
This study makes the following contributions:

\begin{enumerate}
  \item An exploratory empirical characterization of how \keyTerm{lexical bias}, \keyTerm{prompting}, and \keyTerm{external integrations} affect LLM-based vulnerability detection in smart contracts.
  \item A large-scale benchmarking evaluation of 15 LLMs, establishing a realistic baseline for AI-enhanced smart contract auditing.
  \item A methodological contribution in the form of a paired-variant lexical-bias protocol applied at scale, which compares positive and negative variants under matched conditions to isolate superficial lexical cues.
  \item An explicit set of answers to \textbf{RQ1}--\textbf{RQ5}, obtained by comparing matched positive and negative contract variants under controlled prompting and external-tool conditions, and by analyzing the resulting model outputs with our classification pipeline.
  \item An engineering contribution in the form of a reusable reproduction infrastructure and an automatic classifier for unstructured LLM outputs, which achieves \text{92\%} accuracy and enables reproducible large-scale experiments on LLM-based code analysis.
\end{enumerate}

\paragraph*{Paper Organization}
In \Cref{sec:background}, we provide a brief literature review of closely related work. We also present the collection of smart contracts used in our study, along with the main definitions and metrics that we adopt from existing research.
In \Cref{sec:methodology}, we describe our methodology and the infrastructure that we created.
In \Cref{sec:results}, we present the evaluation results that we obtained.
In \Cref{sec:keyfindings}, we highlight our key findings. Finally, in \Cref{sec:conclusion}, we present our conclusions and discuss our future plans.

A shorter version of this paper has been accepted for publication in the Proceedings of the 2026 IEEE International Conference on Computers, Software, and Applications (COMPSAC). This technical report extends the conference version by including the complete experimental result tables, detailed prompt configurations, and extended methodology flowcharts in the appendices.

\section{Preliminaries}
\label{sec:background}
\subsection{Smart Contracts and Blockchain}

Smart contracts~\cite{ethereum} are programs executed within a decentralized environment known as the blockchain. 
These contracts often handle valuable assets, including cryptocurrencies and other types of digital tokens. 
One of the major advantages of smart contracts is their ability to encode agreements that are automatically executed.
The consensus mechanisms implemented by blockchains ensure some degree of immutability, meaning that once deployed, smart contracts cannot be modified or altered. 
This also ensures transparency, as the code is publicly available for anyone, making all actions traceable.

However, this immutability comes with a significant downside: it is extremely risky to deploy buggy or vulnerable contracts. A vulnerability in a smart contract could lead to serious financial and operational consequences. Additionally, since the code is publicly available, it is more exposed than traditional software, making it an attractive target for malicious actors.
Some examples of such vulnerabilities and their consequences can be found in~\cite{8733785,dao,parity,10.1145/3282373.3282419,10.1007/978-3-662-54455-6_8}.
\subsection{Prompting}
Zero-Shot prompting is an interaction paradigm where an LLM is tasked with performing a specific function—such as identifying a smart contract vulnerability—without receiving prior examples or task-specific demonstrations~\cite{Brown2020}.
This technique serves to evaluate the model's inherent ability to generalize its pre-trained knowledge and interpret code logic across diverse software domains~\cite{chen2024unleashingpotentialpromptengineering}.

Chain-of-Thought (CoT) prompting is a strategy designed to elicit logical reasoning by instructing the model to generate intermediate analytical steps before arriving at a final output~\cite{Wei2022}.
By mimicking a human-like, step-by-step deductive process, CoT aims to improve model performance on complex tasks, such as tracing a reentrancy path through inter-contract calls. However, recent findings suggest that while reasoning-intensive approaches effectively filter out superficial noise, they often introduce a trade-off by narrowing the overall discovery range and increasing computational latency~\cite{AppleReasoningModels}.

The interaction with Large Language Models is architecturally segmented into system and user prompts to separate behavioral meta-rules from specific task data.
The system prompt serves as a foundational configuration layer, defining the model's persona, expertise, and persistent operational constraints—such as formatting requirements and tone—following established patterns~\cite{white2023prompt, zhang2024sprigimprovinglargelanguage}. 
In contrast, the user prompt acts as the dynamic input layer, providing the specific, variable content or query intended to drive a particular response for a single request.
By isolating operational directives from task-specific inputs, this dual-layer architecture ensures consistent behavioral alignment and performance across complex applications.

\subsection{LLM Inference Internals}
The Key-Value (KV) cache is a memory optimization mechanism used during the inference of Transformer-based models to store the intermediate mathematical representations of the self-attention layers~\cite{AttentionIsAllYouNeed}.
In autoregressive generation, each new token is generated based on the context of all preceding tokens; the KV cache stores the \say{Keys} and \say{Values} of these past tokens to avoid redundant computations, enabling the model to process only the most recent token at each step~\cite{Pope2023}. 
However, as the cache size increases linearly with sequence length, it can place substantial pressure on the system's video memory (VRAM), potentially leading to performance degradation or hardware failures during long-context analyses~\cite{LostInTheMiddle}.

\subsection{LLM Inference Hyperparameters}
Temperature is a hyperparameter utilized during Large Language Model inference to adjust the randomness or \say{creativity} of the generated output.
It functions by scaling the probability distribution of the next-token prediction; a value of zero makes the model deterministic by ensuring it consistently selects the token with the highest probability, whereas higher values encourage the selection of less likely tokens to produce more diverse responses ~\cite{achiam2023gpt}. 
This parameter is critical for balancing the model's output between strict adherence to learned patterns and exploratory generation.

Structured output refers to the enforcement of a rigid, predefined data format—typically JSON—on the response generated by a Large Language Model to ensure its compatibility with downstream software systems. 
This mechanism is designed to eliminate non-essential conversational text and ensure that all findings follow a strict schema, which is essential for the reliability of automated parsing and classification pipelines. 
LLM providers facilitate this through two methods: request parameterization (schema enforced natively at inference) and instructional enforcement (system prompt constrains the model's linguistic behavior)~\cite{white2023prompt}. 
By converting probabilistic text into deterministic data structures, structured output allows generative models to function as reliable components within complex, programmatic workflows.  
\subsection{Embeddings and Semantic Similarity}
Embeddings are dense, high-dimensional vector representations of textual data where words, phrases, or code snippets are mapped into a continuous mathematical space~\cite{mikolov2013efficientestimationwordrepresentations, pennington2014glove}. 
This transformation allows the model to capture deep semantic relationships, as tokens with similar meanings or functional roles are positioned in close proximity to one another~\cite{AttentionIsAllYouNeed}. 

Semantic similarity is a measurement of the conceptual correspondence between data objects—such as words, phrases, or code snippets—within a high-dimensional mathematical space~\cite{chandrasekaran2021evolution}. 
This approach moves beyond simple lexical or character-level matching by evaluating the proximity of data points based on their underlying meaning as represented by embedding vectors. 
Unlike discrete representations, these embeddings enable vector-space operations to identify patterns and quantify conceptual similarity (e.g., cosine similarity or Euclidean distance).
\subsection{Model Compression}
Model quantization is a compression technique that reduces the numerical precision of a model's weights and activations—typically from 16-bit or 32-bit floating-point numbers to lower-bit formats like 4-bit or 8-bit~\cite{Dettmers2022, Gholami2021}. 
By using fewer bits to represent each parameter, this process significantly decreases the overall memory footprint and VRAM requirements needed for inference~\cite{Frantar2023}. 
While effective for optimizing deployment on limited hardware, the reduction in mathematical resolution can introduce quantization-induced logic loss, potentially impairing the model's capacity for complex logical reasoning~\cite{Dettmers2023}.


\subsection{Related Work}
The consequences of vulnerable smart contracts code have motivated the research community and practitioners to develop detection tools and collections of vulnerable smart contracts.
Currently, there are many such collections~\cite{swc,dasp,smartbugs,defihacks,mabsandmubs,benchmarktests,solidifibenchmark,smartbugscurated,smartbugswild,DBLP:journals/ese/ArusoaieS24}, most of which are written in Solidity~\cite{solidity}, the most popular smart contract language. These collections are organized in different ways, depending on their intended purpose (e.g.,~\cite{swc,dasp,defihacks,smartbugscurated,DBLP:journals/ese/ArusoaieS24}). For example,~\cite{dasp} and~\cite{swc} are community-developed collections primarily aimed at documentation, while also providing a taxonomy of vulnerabilities. In~\cite{defihacks}, the tests are  curated and crafted to be reproducible using Foundry. The collection in~\cite{smartbugscurated} is organized according to the taxonomy from~\cite{dasp}.

\paragraph*{Selected Benchmark}

The collection that we propose in~\cite{DBLP:journals/ese/ArusoaieS24} features a characteristic that enables a more robust evaluation of vulnerability detectors.
This collection follows a comprehensive taxonomy from~\cite{10.3389/fbloc.2022.814977} and is inspired by the Toyota ITC Benchmark~\cite{itc}. The categories of vulnerabilities in this taxonomy are shown in~\Cref{tab:defect_categories}.
The distinctive feature of this collection is its use of pairs of tests that are mostly identical, with one test containing a vulnerability (a.k.a., the \keyTerm{positive} variant) and its counterpart being nearly identical but free from the vulnerability  (a.k.a., the \keyTerm{negative} variant).
In~\cite{DBLP:journals/ese/ArusoaieS24}, the authors exploit this feature to compare existing vulnerability detectors in a robust way: a detector should only signal the vulnerability on the positive variant. A relevant question is whether AI models, which rely on probabilistic methods, can distinguish between the two variants.
Moreover, a comparison of several detectors based on Detection Rate, False Positive Rate, Accuracy, Unique Findings, Precision, Recall, and F-scores is provided. In this work, we use some of these metrics to compare AI models' detections against each other. 

\begin{table}
\caption{Defect categories}\label{tab:defect_categories}
\vspace{-2ex}
\begin{center}
\begin{tabular}{|l|l|}
\hline
\textbf{Category Index} & \textbf{Category Name} \\
\hline
1 & Malicious Environment, Transactions or Input \\
\hline
2 & Blockchain/Environment Dependency \\
\hline
3 & Exception \& Error Handling Disorders \\
\hline
4 & Denial of Service \\
\hline
5  & Resource Consumption \& Gas Issues \\
\hline
6  & Authentication \& Access Control Vulnerabilities \\
\hline
7 & Arithmetic Bugs \\
\hline
8 & Bad Coding and Language Specifics \\
\hline
9 & Environment Configuration Issues \\
\hline
10 & Eliminated/Deprecated Vulnerabilities \\
\hline
\end{tabular}
\end{center}
\vspace{-5ex}
\end{table}

\subsubsection{Similar Approaches}
\label{sec:sim-app}

We highlight two approaches closely related to ours and the differences between our work and theirs.
In~\cite{chen2023chatgpt}, the authors use the GPT ~\cite{achiam2023gpt} and Llama 2~\cite{touvron2023llama}, models available in 2023, to detect vulnerabilities in the Smartbugs curated dataset~\cite{smartbugscurated}. They considered only 7 types of vulnerabilities in order to evaluate how effective they are at detecting vulnerabilities. Their main conclusion is that while ChatGPT demonstrated strengths in pinpointing certain vulnerabilities, it also encountered precision-related challenges. A major difference from their work is that we consider more models and a significantly different benchmark that enables us to identify robust detections.

A key difference in prompt design is that the approach in~\cite{chen2023chatgpt} also includes the expected vulnerability categories.
The prompt we use includes instructions that guide the model towards returning the detections in a specific format, which helps us parse the detections and store them in a database.
The authors of~\cite{chen2023chatgpt} experimented with guiding the models towards a specific format, which reduced the quality of detections. 

Since 2023, there have been great improvements in the capabilities of LLMs to return structured outputs, driven in large part by the wide adoption of Agentic AI tools. Although not true for all the models that we tested, some managed to adhere to the desired response structure for all queries.

The results in~\cite{chen2023chatgpt} show that LLMs achieve a satisfactory recall rate and a large false positive rate. This outcome is similar to the one in our first set of experiments, but in the rest of our experiments, the recall rate dropped significantly. 

Besides measuring the performance of the models, the authors also tested the consistency of responses and identified some causes for the incorrect detections that the LLMs produced.

A more recent study including more models is presented in~\cite{xiao2025logic}. The study included five up-to-date models from different providers: GPT-4o-mini, Gemini 1.5 Pro~\cite{geminiteam2024gemini15unlockingmultimodal}, Claude 3.5 Sonnet, Yi-large~\cite{ai2025yiopenfoundationmodels}, and Qwen-plus.
They used 3 datasets:

\begin{description}[nosep]
    \item [TOP200] consists of open-source, well-audited contract projects primarily used for evaluating FPR.
    \item [Web3Bugs] contains exploitable bugs extracted from audit contests, primarily written in Solidity 0.8, and used to evaluate detection performance across different types of vulnerabilities.
    \item [Messi-Q Smart Contract Dataset] contains data labeled with specific vulnerabilities, written in Solidity v0.4, and used for comparative experiments with older versions.
\end{description}

 One of their most significant findings was the high false positive rate that all models presented, especially Claude 3.5 Sonnet. This finding corresponds with the results obtained in our experiments as well. The authors were able to reduce the number of false detections by designing a prompt that targets one vulnerability at a time. They concluded that having the LLM focus on one task (one vulnerability type) provides a significant decrease in false alarms. 

 Besides the scale of the experiments, our approach distinguishes itself from similar efforts in the following aspects:
 \begin{enumerate}
     \item Our methodology highlights the significant performance volatility that LLMs have depending on code naming conventions, comments, prompting, and external integrations. Thus, quantifying their \keyTerm{lexical bias}, \keyTerm{prompt sensitivity}, and \keyTerm{external data source bias}, concerning aspects regarding the usage of LLMs alone as both standalone detection tools, as well as part of a hybrid system.
     \item Our automatic detection classification technique enabled us to perform multiple large-scale experiments with the possibility of reuse for similar future community efforts.
 \end{enumerate}
 \section{Methodology}
\label{sec:methodology}

\subsection{Strategy}

Our research methodology is structured into four stages: \keyTerm{the preparation of suitable code examples}, \keyTerm{designing the prompts}, \keyTerm{the analysis of the files using our various AI models}, and \keyTerm{the classification of detections in various defect categories}. 
The first two stages were performed manually. The large number of models and vulnerability types led us to implement a reusable framework for automating the last two stages. 

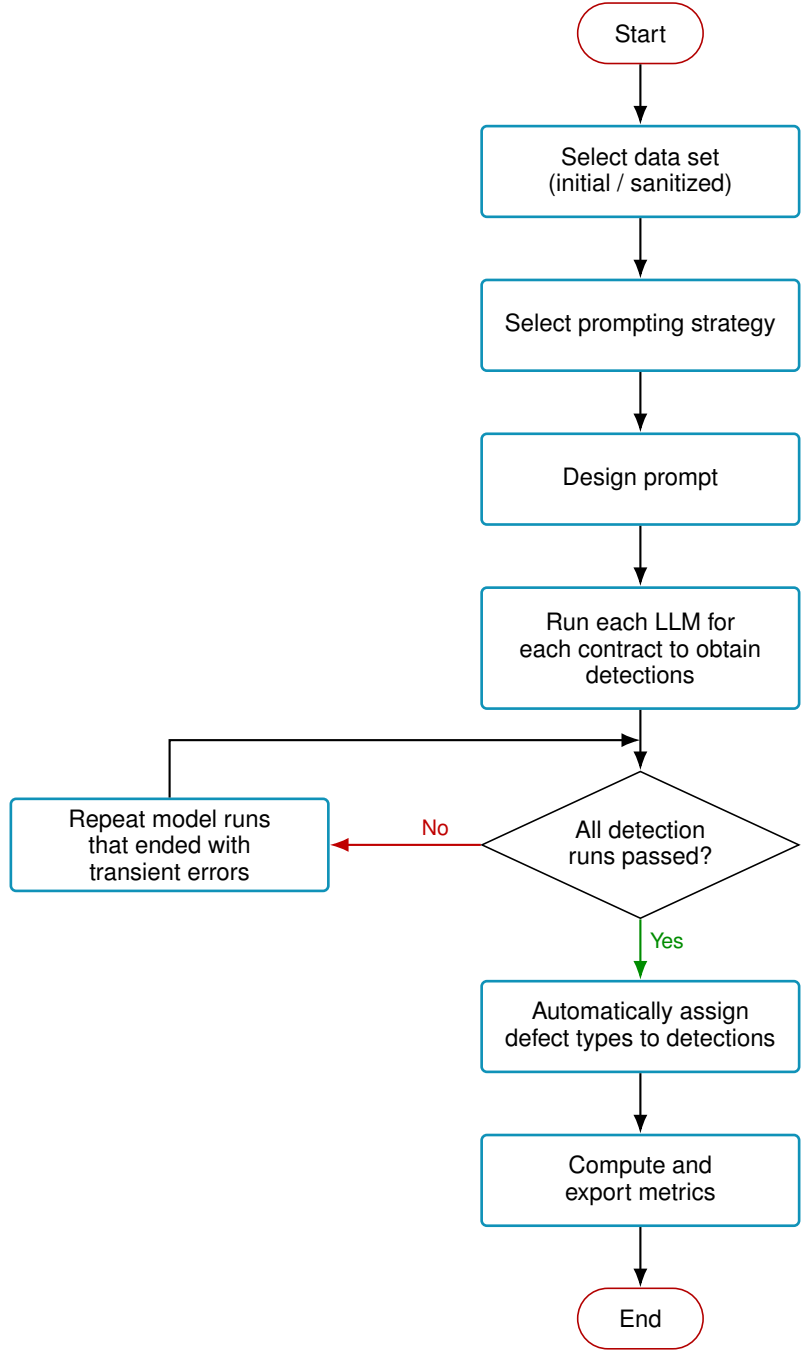
\begin{figure}[H]
  \centering

\begin{tikzpicture}[
    font=\sffamily\small,
    node distance=8mm and 10mm,
    every node/.style={align=center},
    terminator/.style={
        rounded rectangle,
        draw=red!70!black,
        line width=0.6pt,
        minimum width=18mm,
        minimum height=8mm,
        inner sep=2pt
    },
    process/.style={
        rectangle,
        rounded corners=2pt,
        draw=cyan!70!black,
        line width=1pt,
        minimum width=42mm,
        minimum height=12mm,
        inner sep=4pt
    },
    decision/.style={
        diamond,
        aspect=1.6,
        draw=black,
        line width=0.6pt,
        inner sep=1pt,
        minimum width=42mm
    },
    flow/.style={-{Latex[length=2.5mm]}, thick},
    yes/.style={flow, draw=green!55!black},
    no/.style={flow, draw=red!75!black},
]

\node[terminator]                              (start)   {Start};
\node[process,    below=of start]              (sel1)    {Select data set\\(initial / sanitized)};
\node[process,    below=of sel1]               (sel2)    {Select prompting strategy};
\node[process,    below=of sel2]               (design)  {Design prompt};
\node[process,    below=of design,
                  minimum height=16mm]         (run)     {Run each LLM for\\each contract to obtain\\detections};
\node[decision,   below=of run]           (fail)    {All detection \\runs passed?};
\node[process,    left=20mm of fail]
                                               (repeat)  {Repeat model runs\\that ended with\\ transient
                                               errors};
\node[process,    below=of fail]          (assign)  {Automatically assign\\defect types to detections};
\node[process,    below=of assign]             (metrics) {Compute and\\export metrics};
\node[terminator, below=of metrics]            (end)     {End};

\draw[flow] (start)   -- (sel1);
\draw[flow] (sel1)    -- (sel2);
\draw[flow] (sel2)    -- (design);
\draw[flow] (design)  -- (run);
\draw[flow] (run)     -- (fail);

\draw[no]  (fail.west) -- ++(-0.6,0) |- (repeat.east)
            node[pos=0.25, above, font=\sffamily\footnotesize, 
            text=red!75!black] {No};

\draw[yes]   (fail.south) -- (assign.north)
            node[pos=0.35, right, font=\sffamily\footnotesize, text=green!55!black] {Yes};

\draw[flow] (repeat.north) |- ($(run.south)!0.5!(fail.north)$);

\draw[flow] (assign)  -- (metrics);
\draw[flow] (metrics) -- (end);

\end{tikzpicture}
  
  \caption{Methodology flowchart for the automated benchmarking framework.}
  \label{fig:methodology-flowchart}
\end{figure}

The methodology flowchart is provided in \Cref{app:methodology-flowchart} (\Cref{fig:methodology-flowchart}).
The study focuses on Ethereum smart contracts, so we chose Solidity, the language that dominates the deployed smart-contract market and is the only language with paired positive-negative data sets. This choice also reflects the blockchain-specific and language-specific nature of our study and of the benchmark, which are tailored to the execution model, semantics, and tooling ecosystem of smart contracts. In addition, because of its popularity, Solidity is likely to be more frequently represented in LLM training data sets than other smart-contract languages such as Vyper. 
The methodology transfers in principle to
other smart-contract languages. The closest extension is Vyper since it is EVM-compatible and has similar ASTs. Rust-based chains would require a new paired benchmark, and Move would invalidate several defect categories under its resource semantics. Producing such benchmarks is left to future work.
\subsubsection*{Stage 1, preprocessing the collection of defect examples}

We have chosen the test suite presented in~\cite{DBLP:journals/ese/ArusoaieS24} due to its wide variety of defect categories and its positive/negative pairs for each vulnerability. 

This comprehensive range of defect categories allows us to test the capabilities of LLMs in various scenarios. The negative variants also enable us to test the tendency of models to return false alarms.

In addition, the tests in the selected suite are annotated. The defect contained in each code file (for positive variants) or absent from each file (for negative variants) is clearly indicated in the file name or the name of the vulnerable function. 
This feature facilitates metric computation. However, it may also bias AI models toward a specific defect type, since annotated defect category names can act as cues. As a result, such detections may not be fair, because these annotations are not present in practice.
To avoid this, we performed some additional preprocessing of the dataset, such as renaming the contracts, methods, parameters, or variables. 
This allows us to measure the lexical bias.

The original suite contains \textbf{208} code examples distributed across \textbf{10} generic defect categories and \textbf{52} specific vulnerability types. To accommodate our automated classification pipeline, we excluded all examples from \textbf{3} specific categories (see details in \Cref{subsec:automatic_classification}).
This filtering resulted in the final dataset for our study, comprising \textbf{176} code examples across \textbf{49} specific defect categories.

\subsubsection*{Stage 2, designing the prompts}
We designed two prompts. The first requests an analysis report from the model without any additional instructions besides asking it to assume a \say{senior Smart Contract Security Auditor specializing in Solidity} role. For the second, we present the LLM with a structured approach with additional instructions that ask for additional internal validation of the findings. In prompt engineering~(see, e.g,~\cite{chen2024unleashingpotentialpromptengineering,white2023prompt}), these two prompts correspond to \keyTerm{Zero-Shot} and \keyTerm{CoT} prompting. 
We focus on these two canonical prompting families, and we exclude few-shot prompting because per-vulnerability exemplars would leak defect-type cues.
For each of these initial prompts, we generated a variant in which we provided the model with the security report generated by Slither. For the CoT variant, we added an additional step in which we instruct the model to attempt to disprove the detections provided by Slither in an effort to mitigate potential biases. The resulting prompts are used to perform several experiments detailed in \Cref{sec:results}. The complete prompts, including the structured output prompt used as a system prompt for models without dedicated schema-enforcement request parameters, are reproduced in \Cref{app:full-prompts}.

\subsubsection*{Stage 3, analyze defect examples}

We used official client libraries to communicate with both remote and local model endpoints and prompted the models to return structured detections: defect type, general defect definition, and how the defect manifests in the contract being analyzed. The resulting analysis report is then stored in a database. To handle transient API errors, such as server overload or rate limiting, failed analysis runs were stored and subsequently retried. This ensures complete data collection and prevents temporary issues from being incorrectly marked as undetected vulnerabilities. 

\subsubsection*{Stage 4, classify the results of the analysis}

This stage begins by automatically assigning specific defect types to the detections generated during the processing stage. 
Due to the large number of detections that we obtained, manually classifying each detection is not feasible. 
We describe the process of designing and implementing our automatic classification step in greater detail in \Cref{subsec:automatic_classification}. 
Following the assignment of defect types, we assign labels to each output to indicate whether it is a \keyTerm{true positive} (TP), \keyTerm{true negative} (TN), \keyTerm{false positive} (FP), or \keyTerm{false negative} (FN).
Then, relevant performance metrics are computed based on the detections and their assigned labels. We also measure the number of instances where an LLM failed to produce structured output and the number of detections that match the ones provided by Slither.
 
\subsection{Interacting with LLMs}

To ensure that LLM outputs were easy to parse, we asked for a strict JSON-structured format. 
Depending on the provider, we did this using one of the methods below:

\begin{enumerate}
    \item Request Parameterization: For models and providers supporting native schema enforcement (e.g., Grok, Qwen), we integrated the desired response structure into the request parameters to ensure 100\% format compliance.
    \item System Prompt Enforcement: For models lacking native parameter-based JSON support (e.g., Mistral, DeepSeek), we developed a specialized prompt to  enforce the schema and suppress non-essential conversational text.
\end{enumerate}

Runtime and formatting benchmarks were set locally using the Ollama framework. The main experimental pipeline used official developer APIs to mitigate hardware-induced performance bottlenecks and provide access to the full operational capabilities of the models. An exception was made for GPT-Oss 120b, which was accessed via OpenRouter due to its unavailability through the main OpenAI API. We set a timeout limit of \textbf{5} minutes for both local and API-based model queries.

\subsection{Automated Benchmarking Framework}

To operationalize the methodology described above, we implemented a reusable benchmarking framework that minimizes manual intervention during experiment execution and result collection. The framework was designed around four requirements.

\begin{description}
  \item[Automation.] It executes complete experiment batches end to end, including prompt expansion, model invocation, structured-output validation, persistence of intermediate results, and retry handling for transient failures.
  \item[Configurability.] It supports parameterized prompt templates in which system prompts and user prompts can be combined in the same experiment, replacement tokens can be defined for contract code and auxiliary inputs, and individual models can be included or excluded without modifying the core execution flow.
  \item[Model compatibility.] It abstracts remote APIs and local inference backends through a shared execution layer. When a provider requires a dedicated SDK, the framework adapts that SDK to the common interface so that model integration remains uniform.
  \item[Scalability.] Our initial sequential implementation required approximately \textbf{18} hours to complete a full experiment. By batching API-based models by provider and executing independent jobs concurrently, we reduced the runtime to approximately \textbf{10} hours. On the available hardware, however, parallel execution of multiple local models was not feasible because of memory pressure and inference contention, so local runs remained serialized.
\end{description}

\subsection{Automatic defect detection classification}
\label{subsec:automatic_classification}

The approach we propose is grounded in the principles of \keyTerm{Semantic Similarity}~\cite{chandrasekaran2021evolution}, which facilitates the measurement of correspondence between data objects within a high-dimensional conceptual space. This allows for a robust classification of model outputs that goes beyond lexical-based comparison.
We assigned a definition to each defect. To avoid noise, the definitions were concise, consisting of a few phrases. For example, the category \defectCategory{Frozen Ether}
is described as:
\begin{quote}
\textit{
A permanent loss scenario where ether is locked in a contract due to missing withdrawal mechanisms, unreachable conditions (e.g., broken access control), or logic errors preventing legitimate users from retrieving funds.
} 
\end{quote}

We measured the accuracy of our classifier by manually classifying the results from 6 models (approx. 4000 detections) and using them as a reference.
During the manual classification process, we inspected every detection and checked if the defect category and the description of the vulnerability in the context of the contract matched a category present in the taxonomy that our benchmark is based on.

After experimenting with multiple embedding models such as \keyTerm{nomic-embed-text}, \keyTerm{bge-large3}, \keyTerm{granite-embedding}, we obtained the best accuracy using \keyTerm{qwen3-embedding:8b}.

Our automatic similarity-based classifier works as follows.

Initially, we compute the similarity between the defect type name from our list to the defect type name from the detection. If the similarity threshold required for an assignment is not met, then we also perform the same similarity computation for both the defect and its general definition.

We noticed that, in instances where certain defect categories were misclassified, the assigned category closely matched the expected category.
 
Usually, they were different instances of the same general defect. For example, detections for \defectCategory{Gas Costly Loops} were often wrongly assigned as \defectCategory{Gas Costly Pattern}. To increase the effectiveness of our process, we decided to allow it to accept close matches for similar defects. The resulting close-match pairs are listed in \Cref{app:close-match-pairs}, \Cref{tab:close-match-pairs}.

Since we know the defect corresponding to each positive and negative variation, we are assigning that defect to the detection if the classified defect is an exact or a very close match. Otherwise, they remain unassigned to not include out-of-scope detections. By relaxing the assignment constraints in this way, we were able to increase the overall accuracy to over \textbf{80\%} when compared to the results of manual classification. 

However, detections from the defect categories \defectCategory{Unexpected Throw and Revert (3B)}, and \defectCategory{Assert, Require or Revert Violation (3D)} were still not being properly classified. Due to their relatively large number of samples and a detection classification accuracy of less than \textbf{40\%}, we decided to exclude samples from these categories from our study. Even though the sample size for \defectCategory{Callstack Depth Limit (10C)} was small compared to others, we also excluded it due to its low classification accuracy score. This decision increased the accuracy of our automatic detection classification process to \textbf{88\%}. The test suite is organized in categories and subcategories. The defect classes mentioned above are subcategories. Thus, we preserved \textbf{49} out of the \textbf{52} initial subcategories, still having examples from all \textbf{10} larger categories; the retained subcategories are listed in \Cref{app:defect-subcategories}.

We observed that the inherent granularity of defect taxonomies significantly influenced classification accuracy. Specifically, certain categories, such as \textbf{8L} (Inadequate Logging or Documentation), function as broad umbrellas covering diverse issues like missing event emissions or unindexed addresses. In several instances, the models correctly identified a specific defect but failed to map it to the broader parent category, returning the granular issue instead of the generic label. To address this semantic granularity gap, we expanded the classifier's mapping logic by introducing additional synonymous definitions and sub-category identifiers. This refinement better aligned the model's specific detections with the study's general taxonomy, ultimately increasing the classifier's accuracy from \textbf{88\%} to \textbf{92\%}.

\section{Evaluation Process}
\label{sec:results}
\subsection{Overview}
\label{subsec:overview}

Given the description of \textit{Stage 2} in our methodology, we evaluate how each combination performs on our dataset: Zero-Shot with and without Slither input, and CoT with and without Slither input. As described in \textit{Stage~1}, we preprocessed the dataset in~\cite{DBLP:journals/ese/ArusoaieS24} to eliminate the most basic forms of lexical bias. 
To differentiate between the original, unprocessed dataset from~\cite{DBLP:journals/ese/ArusoaieS24} and the processed one used here, we refer to them as Dataset 1 and Dataset 2, respectively. The experiments that we perform and their objectives are shown in~\Cref{tab:experiment_setup}.

\begin{table}[ht]
\centering
\caption{Experimental Design and Objectives\label{tab:experiment_setup}}
\begin{tabular}{|l|l|l|l|l|}
\hline
\textbf{Exp.} & \textbf{Dataset} & \textbf{Prompting} & \textbf{Slither} & \textbf{Primary Objective} \\
\hline
1 & Dataset 1 & Zero-Shot & No & Establish baseline \\
\hline
2 & Dataset 2 & Zero-Shot & No & Quantify Lexical Bias \\
\hline
3 & Dataset 2 & CoT & No & Evaluate reasoning benefit \\
\hline
4 & Dataset 2 & Zero-Shot & Yes & Hybrid tool integration \\
\hline
5 & Dataset 2 & CoT & Yes & Measure Slither Bias \\
\hline
\end{tabular}
\end{table}

\begin{table}
\centering
\caption{Models and their developers included in our study\label{tab:models_developers}}
\begin{tabular}{|l|l|}
\hline
\textbf{Developers} &  \textbf{Models} \\
\hline
DeepSeek &  DeepSeek Chat, DeepSeek Reasoner,\\ & DeepSeek R1 14b\\
\hline
Mistral & Magistral Medium, Magistral 24b (local and API),\\ & Devstral, Devstral Small (local and API)\\
\hline
xAI & Grok 4.1 fast reasoning, Grok 4.1 fast non-reasoning\\
\hline
OpenAI & GPT Oss 120b, GPT Oss 20b (local and API)\\
\hline
Alibaba Cloud & Qwen3 30b (thinking), Qwen3 30b (instruct),\\ & Qwen3 235b (thinking), Qwen3 235b (instruct)\\
\hline
\end{tabular}
\end{table}

\subsection{Setup}

All experiments were performed on a single machine with an M5 chip and 24GB shared RAM/VRAM memory. All local models were run using 4-bit quantization to reduce their memory footprint by decreasing the precision of their weights. No fine-tuning was performed to alter the performance of these local models. The \keyTerm{temperature} parameter was set to \textbf{0}  and the \keyTerm{timeout} for each LLM query was set to \keyTerm{5 minutes}.

\subsection{Model Selection}

The primary criterion for model selection was the feasibility of long-term integration into custom static analysis workflows. To keep reasonable financial costs, we fixed a strict pricing ceiling of \$5 per 1 million input/output tokens. This threshold implied the exclusion of flagship models from numerous providers, such as Claude 4.5 Opus (Anthropic), which we categorize as too costly for large-scale auditing tasks.

Following this financial filtering, we selected models specifically to address the objectives of model scaling and architectural comparison. We prioritized models that exist in paired series, allowing for direct comparisons regarding model size and reasoning capabilities. Models that met the pricing criteria but lacked a relevant counterpart for scaling or reasoning analysis were excluded to maintain the thematic consistency of the experimental matrix.

The final list of included models is shown in~\Cref{tab:models_developers}, and the API providers used to access them are listed in \Cref{app:api-providers}, \Cref{tab:api-providers}.

\subsection{Metrics}
We adopt the performance metrics defined in~\cite{DBLP:journals/ese/ArusoaieS24}.
The presence of positive/negative variations within the test suite is essential: we compute the False Positive Rate (FPR) and other metrics that quantify the \say{noise} inherent in LLM outputs. 

We only show here \keyTerm{Recall} and \keyTerm{FPR}.
Based on these, we compute the Balanced Accuracy (BA) to enable objective comparison across our scenarios and address our research questions.

\subsubsection{Recall} Measures the proportion of actual positive instances that were correctly identified by the model: 
$$\mathit{Recall} = {\it TP}/({\it TP + FN}),$$
where {\it TP} is the number of true positives and {\it FN} is the number of false negatives. Recall is particularly important when the cost of missing a positive instance (a false negative) is high.

\subsubsection{False Positive Rate ($\mathit{FPR}$)}
Represents the proportion of negative instances that are incorrectly classified as positive:
$$\mathit{FPR} = {\it FP}/({\it FP + TN}),$$
where {\it FP} is the number of false positives and {\it TN} is the number of true negatives. A lower value for $\mathit{FPR}$ is better.

\subsubsection{Balanced Accuracy}
It is computed as 
\[\mathit{BA}=\big({\mathit{Recall} + (1 - \mathit{FPR})}\big)/2,\]
\noindent
and tells us how well a tool finds positives and avoids false alarms. A higher value for $\mathit{BA}$ is better.
\subsection{Results}
General experiment-level results (detection metrics,
confusion-matrix counts, and reliability per experiment) are summarized in \Cref{app:experiment-results}, while the per-category results are available in the companion repository referenced in \Cref{sec:data_availability}.

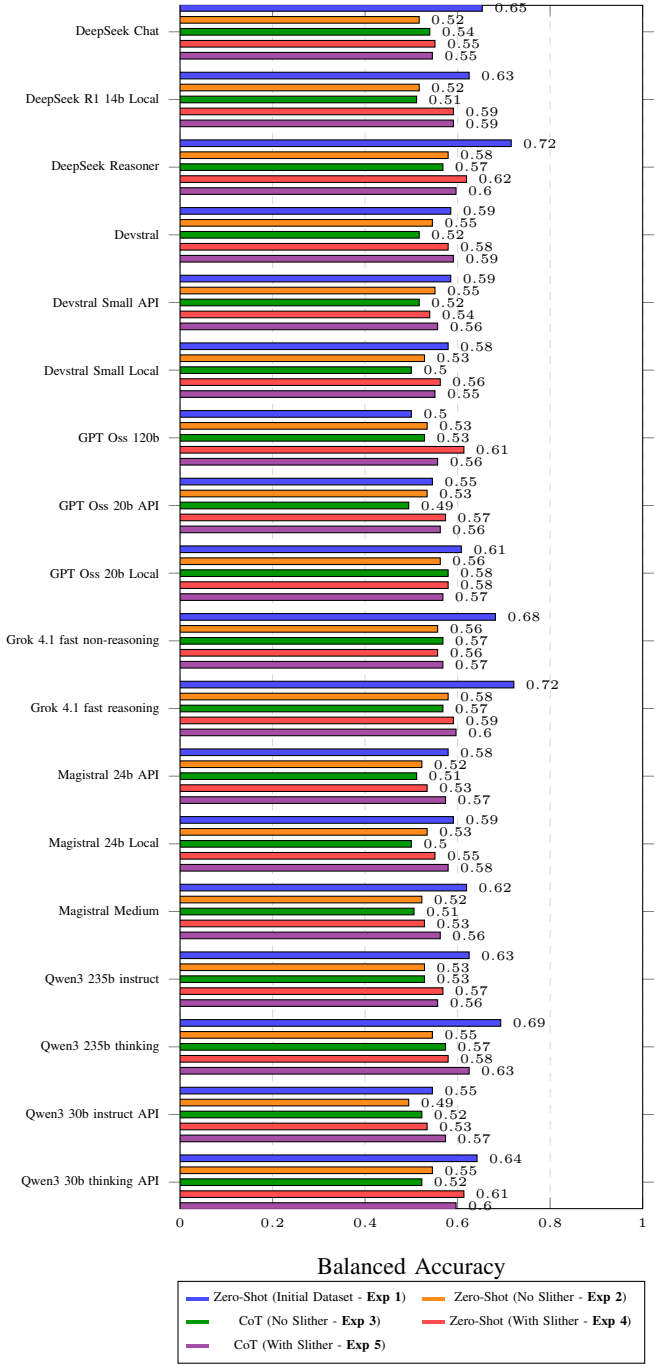
\begin{figure}[t]
  \begin{tikzpicture}
  
    \begin{axis}[
      xbar,
      width=\columnwidth/1.15,
      height=17.5cm,
      bar width=2.5pt,
      enlarge y limits={abs=0.35cm},
      xlabel={Balanced Accuracy},
      ylabel={},
      symbolic y coords={Qwen3 30b thinking API, Qwen3 30b instruct API, Qwen3 235b thinking, Qwen3 235b instruct, Magistral Medium, Magistral 24b Local, Magistral 24b API, Grok 4.1 fast reasoning, Grok 4.1 fast non-reasoning, GPT Oss 20b Local, GPT Oss 20b API, GPT Oss 120b, Devstral Small Local, Devstral Small API, Devstral, DeepSeek Reasoner, DeepSeek R1 14b Local, DeepSeek Chat},
      ytick=data,
      yticklabel style={font=\tiny},
      xticklabel style={font=\tiny},
      xlabel style={font=\small},
      xmin=0, xmax=1,
      legend style={
        font=\tiny,
        at={(0.5,-0.06)},
        anchor=north,
        legend columns=2,
        /tikz/every even column/.append style={column sep=4pt},
      },
      legend image code/.code={
        \draw[#1, draw=none] (0cm,-0.02cm) rectangle (0.3cm,0.02cm);
      },
      nodes near coords,
      nodes near coords style={font=\fontsize{3}{4}\selectfont, anchor=west},
      every node near coord/.append style={xshift=1pt},
      xmajorgrids=true,
      grid style={dashed, gray!30},
      reverse legend,
    ]

    \addplot[fill=violet!70] coordinates {(0.5455,DeepSeek Chat) (0.5909,DeepSeek R1 14b Local) (0.5966,DeepSeek Reasoner) (0.5909,Devstral) (0.5568,Devstral Small API) (0.5511,Devstral Small Local) (0.5568,GPT Oss 120b) (0.5625,GPT Oss 20b API) (0.5682,GPT Oss 20b Local) (0.5682,Grok 4.1 fast non-reasoning) (0.5966,Grok 4.1 fast reasoning) (0.5739,Magistral 24b API) (0.5795,Magistral 24b Local) (0.5625,Magistral Medium) (0.5568,Qwen3 235b instruct) (0.625,Qwen3 235b thinking) (0.5739,Qwen3 30b instruct API) (0.5966,Qwen3 30b thinking API)};
    \addlegendentry{CoT (With Slither - \textbf{Exp 5})}

    \addplot[fill=red!70] coordinates {(0.5511,DeepSeek Chat) (0.5909,DeepSeek R1 14b Local) (0.6193,DeepSeek Reasoner) (0.5795,Devstral) (0.5398,Devstral Small API) (0.5625,Devstral Small Local) (0.6136,GPT Oss 120b) (0.5739,GPT Oss 20b API) (0.5795,GPT Oss 20b Local) (0.5568,Grok 4.1 fast non-reasoning) (0.5909,Grok 4.1 fast reasoning) (0.5341,Magistral 24b API) (0.5511,Magistral 24b Local) (0.5284,Magistral Medium) (0.5682,Qwen3 235b instruct) (0.5795,Qwen3 235b thinking) (0.5341,Qwen3 30b instruct API) (0.6136,Qwen3 30b thinking API)};
    \addlegendentry{Zero-Shot (With Slither - \textbf{Exp 4})}

    \addplot[fill=green!60!black] coordinates {(0.5398,DeepSeek Chat) (0.5114,DeepSeek R1 14b Local) (0.5682,DeepSeek Reasoner) (0.517,Devstral) (0.517,Devstral Small API) (0.5,Devstral Small Local) (0.5284,GPT Oss 120b) (0.4943,GPT Oss 20b API) (0.5795,GPT Oss 20b Local) (0.5682,Grok 4.1 fast non-reasoning) (0.5682,Grok 4.1 fast reasoning) (0.5114,Magistral 24b API) (0.5,Magistral 24b Local) (0.5057,Magistral Medium) (0.5284,Qwen3 235b instruct) (0.5739,Qwen3 235b thinking) (0.5227,Qwen3 30b instruct API) (0.5227,Qwen3 30b thinking API)};
    \addlegendentry{CoT (No Slither - \textbf{Exp 3})}

    \addplot[fill=orange!90] coordinates {(0.517,DeepSeek Chat) (0.517,DeepSeek R1 14b Local) (0.5795,DeepSeek Reasoner) (0.5455,Devstral) (0.5511,Devstral Small API) (0.5284,Devstral Small Local) (0.5341,GPT Oss 120b) (0.5341,GPT Oss 20b API) (0.5625,GPT Oss 20b Local) (0.5568,Grok 4.1 fast non-reasoning) (0.5795,Grok 4.1 fast reasoning) (0.5227,Magistral 24b API) (0.5341,Magistral 24b Local) (0.5227,Magistral Medium) (0.5284,Qwen3 235b instruct) (0.5455,Qwen3 235b thinking) (0.4943,Qwen3 30b instruct API) (0.5455,Qwen3 30b thinking API)};
    \addlegendentry{Zero-Shot (No Slither - \textbf{Exp 2})}
    
    \addplot[fill=blue!70] coordinates {(0.6534,DeepSeek Chat) (0.625,DeepSeek R1 14b Local) (0.7159,DeepSeek Reasoner) (0.5852,Devstral) (0.5852,Devstral Small API) (0.5795,Devstral Small Local) (0.5,GPT Oss 120b) (0.5455,GPT Oss 20b API) (0.608,GPT Oss 20b Local) (0.6818,Grok 4.1 fast non-reasoning) (0.7216,Grok 4.1 fast reasoning) (0.5795,Magistral 24b API) (0.5909,Magistral 24b Local) (0.6193,Magistral Medium) (0.625,Qwen3 235b instruct) (0.6932,Qwen3 235b thinking) (0.5455,Qwen3 30b instruct API) (0.642,Qwen3 30b thinking API)};
    \addlegendentry{Zero-Shot (Initial Dataset - \textbf{Exp 1})}

    \end{axis}
    
  \end{tikzpicture}
  \caption{Balanced accuracy across all experiments.\label{fig:balanced_accuracy}}
\end{figure}

\subsection{Analysis of the Results}

Our analysis of the results addresses each research question, with aggregated changes in balanced accuracy reported.

\subsubsection{Quantifying the lexical bias between the two datasets}
To answer \textbf{RQ1}, we compare Experiment 1 (which uses the original defect examples containing syntactical clues that LLMs might exploit) against all other experiments (which use the preprocessed defects suite with syntactical clues removed).

The raw data shows that Experiment 1 consistently achieves higher BA than Experiments 2–5 (cf.~\Cref{tab:exp1vsall}), indicating the presence of \keyTerm{lexical bias}. This can be seen in~\Cref{fig:balanced_accuracy}, where the Zero-Shot blue bars have better scores. This suggests that specific code wording may enable LLMs to detect defects through lexical cues rather than by reasoning about the underlying issue.

\begin{table}
\centering
\caption{Comparison of BA between Experiments 2–5 and Experiment 1. Values indicate lower detection performance in Experiments 2–5 compared to Experiment 1, suggesting that in Experiment 1 the superior accuracy is due to lexical bias.\label{tab:exp1vsall}}
\begin{tabular}{@{}lllrl@{}}
\hline
Exp $i$ vs. Exp $j$ & Result & $\Delta$BA & Status \\
\hline
Exp 2 vs. Exp 1 & WORSE & -0.077652 & \textcolor{red!60}{\rule{2.33cm}{0.3cm}} \\
Exp 3 vs. Exp 1 & WORSE & -0.085543 & \textcolor{red!60}{\rule{2.57cm}{0.3cm}} \\
Exp 4 vs. Exp 1 & WORSE & -0.046086 & \textcolor{red!60}{\rule{1.38cm}{0.3cm}} \\
Exp 5 vs. Exp 1 & WORSE & -0.041351 & \textcolor{red!60}{\rule{1.24cm}{0.3cm}} \\
\hline
\end{tabular}
\end{table}

\subsubsection{Analyzing the impact of using CoT prompting}

\begin{table}
\centering
\caption{Comparison of BA between experiments with Zero-Shot and CoT. Values do not indicate an improvement when using CoT.\label{tab:cotvszeroshot}}
\begin{tabular}{@{}lllrl@{}}
\hline
Exp $i$ vs. Exp $j$ & Result & $\Delta$BA & Status \\
\hline
Exp 3 vs. Exp 2 & WORSE & -0.007891 & \textcolor{red!60}{\rule{0.24cm}{0.3cm}} \\
Exp 5 vs. Exp 4 & BETTER & 0.004735 & \hfill\textcolor{green!60}{\rule{0.14cm}{0.3cm}} \\
\hline
\end{tabular}
\end{table}

To answer \textbf{RQ2}, we compare Experiments 2 and 3 (Zero-Shot vs. CoT without Slither) and Experiments 4 and 5 (Zero-Shot vs. CoT with Slither) to assess whether CoT outperforms Zero-Shot. \Cref{tab:cotvszeroshot} shows that without Slither, CoT performs slightly worse, while with Slither, CoT performs slightly better. 

However, on aggregate, the $\Delta$BA--defined as the numerical difference in BA between two experimental configurations--between CoT and Zero-Shot is within per-model variance ($\pm 0.06$).

Notably, the effect of CoT varies across models: for Qwen3 30b instruct API, Qwen3 235b thinking, and Magistral 24b API, CoT improves BA by approximately 0.04, while for other models (e.g., GPT Oss 120b) it decreases BA by up to 0.06.

In addition to the above, CoT increased the computational overhead, as models were required to generate an internal chain of reasoning. 
The time that was necessary for the models to respond increased by up to \textbf{10} seconds. Almost all models queried via API did not suffer shifts in structured output consistency compared to Experiment 2. However, a general increase in invalid outputs was noticed for the local models, the number of such instances increasing by between \textbf{10} and \textbf{20}, effectively doubling for most models.

\subsubsection{Assessing the performance induced by integrating LLMs with Slither}
An answer for \textbf{RQ3} requires us to evaluate the performance boost introduced when LLMs are supplemented with external security reports, such as those from Slither. Therefore, we compare Zero-Shot with/without Slither (Exp 2 vs. Exp 4) and CoT with/without Slither (Exp 3 vs. Exp 5). 

The results are shown in~\Cref{tab:withslither}.
The results indicate that incorporating Slither input significantly improves accuracy. 

\begin{table}
\centering
\caption{Comparison of BA between experiments with/without Slither input. Values for CoT experiments 4 and 5 do indicate an improvement when using Slither input for LLMs.\label{tab:withslither}}
\begin{tabular}{@{}lllrl@{}}
\hline
Exp $i$ vs. Exp $j$ & Result & $\Delta$BA & Status \\
\hline
Exp 4 vs. Exp 2 & BETTER & 0.031566 & \hfill\textcolor{green!60}{\rule{0.95cm}{0.3cm}} \\
Exp 5 vs. Exp 3 & BETTER & 0.044192 & \hfill\textcolor{green!60}{\rule{1.33cm}{0.3cm}} \\
\hline
\end{tabular}
\end{table}

\subsubsection{Assessing the bias induced by grounding LLMs with Slither detections}
\begin{figure}[t]
  \centering
  \begin{tikzpicture}
    \begin{axis}[
      xbar,
      width=\columnwidth/1.15,
      height=16cm,
      bar width=3pt,
      enlarge y limits={abs=0.35cm},
      xlabel={Count of Detections Matching Slither},
      ylabel={},
      symbolic y coords={Qwen3 30b thinking API, Qwen3 30b instruct API, Qwen3 235b thinking, Qwen3 235b instruct, Magistral Medium, Magistral 24b Local, Magistral 24b API, Grok 4.1 fast reasoning, Grok 4.1 fast non-reasoning, GPT Oss 20b Local, GPT Oss 20b API, GPT Oss 120b, Devstral Small Local, Devstral Small API, Devstral, DeepSeek Reasoner, DeepSeek R1 14b Local, DeepSeek Chat},
      ytick=data,
      yticklabel style={font=\tiny},
      xticklabel style={font=\tiny},
      xlabel style={font=\small},
      legend style={
        font=\tiny,
        at={(0.5,-0.08)},
        anchor=north,
        legend columns=2,
        /tikz/every even column/.append style={column sep=4pt},
      },
            legend image code/.code={
        \draw[#1, draw=none] (0cm,-0.02cm) rectangle (0.3cm,0.02cm);
      },
      nodes near coords,
      nodes near coords style={font=\fontsize{3}{4}\selectfont, anchor=west},
      every node near coord/.append style={xshift=1pt},
      xmajorgrids=true,
      grid style={dashed, gray!30},
      reverse legend,
    ]

            \addplot[fill=red!70] coordinates {(104,DeepSeek Chat) (124,DeepSeek R1 14b Local) (81,DeepSeek Reasoner) (123,Devstral) (127,Devstral Small API) (117,Devstral Small Local) (84,GPT Oss 120b) (74,GPT Oss 20b API) (48,GPT Oss 20b Local) (114,Grok 4.1 fast non-reasoning) (65,Grok 4.1 fast reasoning) (98,Magistral 24b API) (106,Magistral 24b Local) (101,Magistral Medium) (136,Qwen3 235b instruct) (42,Qwen3 235b thinking) (114,Qwen3 30b instruct API) (25,Qwen3 30b thinking API)};
    \addlegendentry{CoT (With Slither - \textbf{Exp 5})}

    \addplot[fill=green!60!black] coordinates {(74,DeepSeek Chat) (33,DeepSeek R1 14b Local) (57,DeepSeek Reasoner) (59,Devstral) (66,Devstral Small API) (55,Devstral Small Local) (16,GPT Oss 120b) (20,GPT Oss 20b API) (15,GPT Oss 20b Local) (73,Grok 4.1 fast non-reasoning) (40,Grok 4.1 fast reasoning) (43,Magistral 24b API) (59,Magistral 24b Local) (57,Magistral Medium) (75,Qwen3 235b instruct) (41,Qwen3 235b thinking) (65,Qwen3 30b instruct API) (17,Qwen3 30b thinking API)};
    \addlegendentry{CoT (No Slither - \textbf{Exp 3})}

        \addplot[fill=orange!90] coordinates {(129,DeepSeek Chat) (132,DeepSeek R1 14b Local) (122,DeepSeek Reasoner) (152,Devstral) (134,Devstral Small API) (140,Devstral Small Local) (118,GPT Oss 120b) (104,GPT Oss 20b API) (102,GPT Oss 20b Local) (164,Grok 4.1 fast non-reasoning) (146,Grok 4.1 fast reasoning) (129,Magistral 24b API) (132,Magistral 24b Local) (126,Magistral Medium) (157,Qwen3 235b instruct) (118,Qwen3 235b thinking) (130,Qwen3 30b instruct API) (127,Qwen3 30b thinking API)};
    \addlegendentry{Zero-Shot (With Slither - \textbf{Exp 4})}
    
      \addplot[fill=blue!70] coordinates {(88,DeepSeek Chat) (50,DeepSeek R1 14b Local) (73,DeepSeek Reasoner) (89,Devstral) (72,Devstral Small API) (70,Devstral Small Local) (50,GPT Oss 120b) (17,GPT Oss 20b API) (27,GPT Oss 20b Local) (78,Grok 4.1 fast non-reasoning) (75,Grok 4.1 fast reasoning) (56,Magistral 24b API) (81,Magistral 24b Local) (76,Magistral Medium) (82,Qwen3 235b instruct) (46,Qwen3 235b thinking) (62,Qwen3 30b instruct API) (28,Qwen3 30b thinking API)};
    \addlegendentry{Zero-Shot (No Slither - \textbf{Exp 2})}

    \end{axis}
  \end{tikzpicture}
  \caption{Slither bias: comparison of matched detections across experiments.
  Bars show the number of LLM detections that match Slither findings for
  Zero-Shot and CoT prompting, with and without Slither context.\label{fig:slither_matched_detections}}
\end{figure}
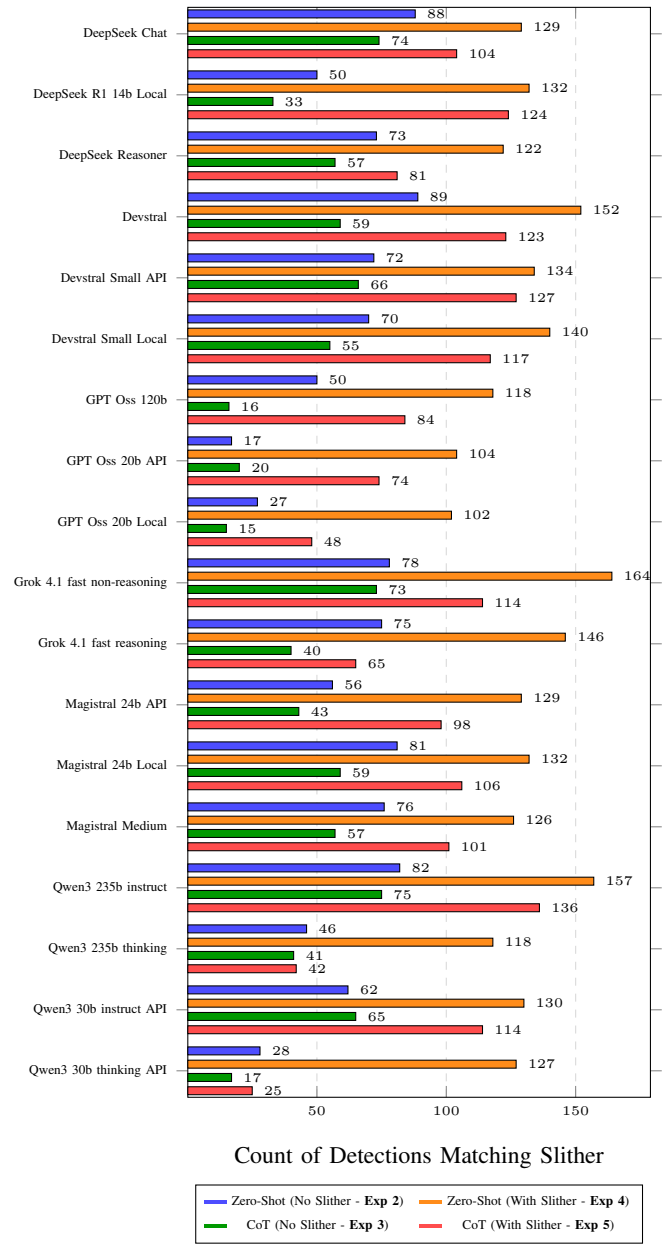

To answer \textbf{RQ4} we need to compare the matched detections between LLMs and Slither. More precisely, we count the number of LLM detections that match Slither detections in all scenarios. Because this analysis differs from the preceding BA-based evaluations, we report the full results in~\Cref{fig:slither_matched_detections}. 
The number of matching detections is higher when Slither is used.

Without an explicit internal validation step, models in Experiment 4 often adopted Slither's findings with limited additional scrutiny, leading to substantial increases in matching detections. This reliance on external data was most evident in models like Qwen3 30b thinking API, which saw a \textbf{354\%} increase in detections matching Slither's report compared to its standalone baseline. This suggests that in Zero-Shot mode, the model functions less as an autonomous auditor and more as a parser for the provided deterministic tool output.

In the CoT experiments, models were instructed to follow a structured approach for validating or challenging Slither's findings through an \keyTerm{adversarial pressure test}. This test required models to determine whether a Slither finding was benign or to identify exploits in functions Slither marked as safe. However, the results indicate that CoT often functioned as a bias amplifier rather than a skeptical filter. Instead of using the reasoning chain to identify false leads, many models used the additional computational steps to generate logical-sounding justifications for the tool's detections to maintain token-to-token coherence. DeepSeek R1 14b Local demonstrated this clearly, with Slither-aligned detections increasing by \textbf{276\%} in the CoT hybrid mode. This failure to maintain an independent adversarial stance suggests that the deterministic signal from a known security tool carries a higher weight in the model's attention mechanism than hypothetical adversarial instructions.

In Zero-Shot, the bias appears consistent with the model's attention mechanism prioritizing structured, deterministic input (Slither reports) over ambiguous raw code. In CoT, the bias appears to stem from instructional alignment; because the model is tasked with analyzing the report, it may treat agreement with the tool as a signal of task completion. This \say{Stochastic Mirror} effect raises a significant concern for agentic AI systems. If an autonomous agent cannot effectively question its toolset or identify semantic vulnerabilities that lie outside rigid rule-sets, it cannot provide the adversarial redundancy required for an effective audit process.

\subsubsection{The feasibility of running models locally}
Because proprietary contracts cannot be sent to third-party APIs under that on-premises constraint, we compare local 4-bit deployment against full-precision API-based models for security auditing.
To answer \textbf{RQ5} we evaluate the architectural and operational trade-offs of deploying LLMs locally using 4-bit quantization vs. using full-precision API-based models. 
Our findings reveal that local deployment is feasible for auditing tasks, but it introduces a class of persistent errors that differentiate them from the transient failures typical of remote services. 
While API-based models occasionally suffered from  issues like server overloads, which can be resolved via automated retries, local models Devstral Small and Magistral 24b exhibited deterministic failures on a recurring subset of contracts that persisted across multiple re-runs. Local failures also manifested as timeouts, driven by the hardware's inability to complete inference within 5-minute timeout window.

Comparing local and API-based models reveals quantization's impact on performance:

\begin{enumerate}
    \item Recall drops: API models consistently outperformed their local counterparts; for instance, the API variant of Devstral Small achieved 55.68\% recall in Experiment 1, while the local variant dropped to 48.86\%.
    \item False Positive Rates: Local models often exhibited higher noise levels, with the local deployment of Magistral 24b showing an FPR of 34.09\% in Experiment 1 compared to the 25.00\% from the API alternative.
    \item Latency Spikes: The average processing time for local models was significantly higher; for example, Magistral 24b required 44624.74 ms per contract on average for Experiment 3, whereas the API variant required only 2\,949.55 ms.
\end{enumerate}

From an architectural standpoint, these discrepancies are largely a consequence of quantization-induced logic loss and hardware constraints. To fit within the 24GB RAM/VRAM limitation of the experimental machine, local models were compressed to 4-bit precision, which can distort weight distributions necessary for tasks like smart contract auditing. This loss causes the model to be unable to handle tasks involving complex logic, leading to the observed consistency in local errors. Furthermore, the hardware faces extreme pressure when managing the KV cache required for CoT reasoning alongside external Slither reports, leading to a high frequency of invalid or truncated JSON outputs. The mentioned issue regarding KV cache and truncated outputs are the causes of persistent errors. Initially, the errors were presented as connection failures to the local Ollama server. In further investigation, we identified these two issues as the causes of the connection shutdown.

\subsection{Addressing the Automatic Classifier Error Rate}
 
To assess the impact of the corresponding \textbf{8\%} error rate of our automatic classifier, we identified its likely failure modes.

Our classifier, which depends on semantic similarity, often fails to recognize true positives that are phrased in atypical ways, resulting in their misclassification as false negatives. This primary error mode would artificially lower the reported Recall, Precision, and Accuracy metrics.
A secondary, less likely error involves the classifier failing to match a FP from an LLM to any known vulnerability, causing it to be misclassified as a TN. This would lead to an underestimation of the FPR.   

With this understanding, we evaluated the robustness of our key findings against a worst-case scenario.
\begin{description}[nosep,leftmargin=0pt]
    \item [RQ1: Lexical Bias] Our finding that \keyTerm{lexical bias} significantly inflates LLM performance remains sound. The transition from the original code (Experiment 1) to the sanitized version (Experiment 2) resulted in a \say{substantial reduction in Recall for every evaluated model}. Because the classifier's primary error (TP $\rightarrow$ FN) deflates Recall, the actual performance drop observed in our experiments is likely even more pronounced than reported, confirming the vulnerability of models to identifier obfuscation.
    \item[RQ2: Reasoning vs. Pattern Matching] Our conclusion that CoT prompting functions primarily as a noise filter is robust. For example, Grok 4.1 fast reasoning showed an improvement in FPR at the cost of Recall. Even if the classifier misclassified some atypical findings, the consistent trend of increased caution and reduced discovery range across models validates that CoT shifts the model towards structured validation rather than broad discovery.
    \item [RQ3: Hybrid Integration Performance] Considering the primary error mode of the classifier, the performance boost recorded when grounding LLMs with Slither reports represents a conservative underestimation of the actual improvement. The distinct upward trend in accuracy remains statistically sound despite this marginal classification noise.
    \item [RQ4: Hybrid Integration Bias] The \say{Stochastic Mirror} effect is not threatened by an 8\% error margin. We observed massive spikes in detections matching Slither—such as the 354\% increase for Qwen3 30b thinking API—and a 276\% increase for DeepSeek R1 14b Local in CoT hybrid mode. Since the secondary failure mode (FP $\rightarrow$ TN) would lead to an underestimation of the FPR, the actual grounding bias and reliance on external tool outputs are likely even higher than our measurements suggest.
    \item[RQ5: Architectural Efficiency] The performance and reliability trade-offs for local 4-bit quantized models are driven by physical hardware constraints and logic loss. The persistent errors identified were primarily connection shutdowns or truncated JSON outputs caused by KV cache pressure. As these are infrastructure-level failures rather than semantic misclassifications, the classifier's error rate does not impact our finding that local deployment introduces a distinct class of non-transient errors.
\end{description}

\section{Key Takeaways}
\label{sec:keyfindings}

This section outlines the key overall findings we observed following our experiments. These conclusions address the potential of LLMs and strategies for using them more effectively.

\subsection{Enhanced Grounding for Hybrid Solutions}
Directly grounding models with security reporting (e.g., Slither) induces a significant grounding bias, providing little benefit over running analysis tools and LLMs independently. Because models cannot currently function as reliable standalone detection tools, external integrations are mandatory; however, providing high-level findings often leads the model to act as a \say{stochastic mirror} rather than a critical auditor.

We propose shifting the grounding layer towards deterministic contract representations, such as the Control Flow Graph (CFG) and Static Single Assignment (SSA) forms. 
While Slither computes this information internally, the current Slither MCP implementation does not expose it through its endpoints. 
Designed mostly for IDE plugins like GitHub Copilot, it currently only exposes limited data, such as function names and access modifiers, which is insufficient for a deeper analysis.

A further avenue for future work is to reposition the LLM as a property-based test generator for formal verification tools such as Kontrol or Forge. 
\subsection{A multi-staged prompting approach}
In addition to the required improvements in the grounding mechanism, our results demonstrate that, for agentic security workflows, a single prompting strategy may not suffice. 
Instead, the most effective systems should adopt a tiered architecture: use Zero-Shot prompting for broad discovery to capture a wide range of potential vulnerabilities, followed by a CoT verification pass to refine the results. 
This allows the system to maintain a high discovery range (Recall) while using CoT to reduce false positives.
By using both approaches and dividing them into different stages, agentic solutions can achieve a more reliable and human-readable audit trail without the performance collapse typically seen in single-pass reasoning.
\subsection{The Quantization Shortcomings on Local Agents}

The performance gap between local 4-bit quantized models and full-precision APIs remains a significant limitation for agentic solutions. While running models locally offers immense financial and privacy advantages, the current trade-off is substantial: local models exhibit persistent, non-transient errors and extreme latency. However, the high cost of accessing top-tier API models remains a practical limitation, making lower-cost models a more viable short-term compromise.

\section{Conclusion and Future Works}
\label{sec:conclusion}
We presented a large-scale evaluation of 15 LLMs and an automated classification pipeline applied to paired positive/negative Solidity examples. The experiments demonstrate four practical conclusions: LLM detections are sensitive to lexical cues; Chain-of-Thought prompting reduces false positives at the cost of recall and latency; grounding with static-analysis reports increases agreement but can introduce grounding bias; and 4-bit local deployments currently suffer quantization-induced reliability issues.

Taken together, these results indicate that LLMs are not yet reliable as standalone security auditors but can provide value within carefully designed hybrid pipelines. Promising directions include grounding models on deterministic program representations (e.g., CFG/SSA) rather than high-level report text; adopting multi-stage pipelines that separate broad discovery from verification; and improving local-model robustness to mitigate quantization loss.
Planned future work includes:
\begin{itemize*}
\item
extending the evaluation to real-world contracts (including deployed samples from ecosystems such as Solana), 
\item
exposing finer-grained representations from Slither (via an MCP extension),  
\item 
exploring property-based test generation for formal verifiers,  
\item
extending the study to other smart-contract languages such as Rust, and  
\item
investigating methods to reduce logic loss in compressed, on-premise models.
\end{itemize*}

\paragraph*{Data Availability}
\label{sec:data_availability}
The datasets, the embeddings used for automatic classification, and the full per-category results are available in the benchmark repository~\cite{benchmark_repo}.

\section*{Acknowledgments}
This work was supported by the \textit{CHIST-ERA-23-SMARTC-01-SCEAL} project, by a grant from the Romanian Ministry of Research, Innovation and Digitization, CNCS/CCCDI - UEFISCDI, project no. 86/2025 ERANET-CHISTERA-IV-SCEAL, within PNCDI IV.

\appendices
\crefalias{section}{appendix}

\onecolumn

\section{Methodology Flowchart}
\label{app:methodology-flowchart}

This diagram provides a compact overview of the end-to-end benchmarking pipeline used in our study, from dataset preparation and prompt configuration to model execution and detection classification. It is intended as a reference map for the experimental workflow discussed in Section~\ref{sec:methodology}.

\section{Prompts Used for LLM Evaluation}
\label{app:full-prompts}

The prompts below are reproduced here and include a short description of the intended use of each prompt. The structured output prompt is the system prompt used to ensure schema-conformant output for models without a dedicated request parameter.

\subsection{Zero-Shot Prompt}

Baseline prompt without external tool grounding.

\begin{quote}
\small\raggedright
**Role:** You are a senior Smart Contract Security Auditor specializing in Solidity.

**Task:** Conduct a deep-dive security analysis of the provided contract(s).

**Code:**

\hspace*{1em}\texttt{\detokenize{[CONTRACT_CODE]}}
\end{quote}

\subsection{Zero-Shot + Slither Prompt}

Baseline prompt with Slither grounding.

\begin{quote}
\small\raggedright
**Role:** You are a senior Smart Contract Security Auditor specializing in Solidity.

**Task:** Conduct a deep-dive security analysis of the provided contract(s).

**Code:**

\hspace*{1em}\texttt{\detokenize{[CONTRACT_CODE]}}

**Slither Output:**

\hspace*{1em}\texttt{\detokenize{[SLITHER_DETECTIONS]}}

Use the Slither output to ground your analysis, but do not limit your findings to these detections; perform a full manual review.
\end{quote}

\subsection{CoT Prompt}

Reasoning-oriented prompt without Slither grounding.

\begin{quote}
\small\raggedright
**Role:** You are a senior Smart Contract Security Auditor specializing in Solidity and EVM-level exploits.

**Task:** Conduct a high-precision security analysis of the provided contract(s).

**Execution Directives:**

1. Invariant Mapping: Identify the critical security invariants of this contract (e.g., "Total deposits must always equal or exceed the sum of individual balances").
2. Adversarial State Analysis: Systematically analyze every state-changing function. Determine if a sequence of transactions--potentially involving multiple users or flash-loan-funded interactions--can violate the identified invariants or bypass access controls.
3. Logic-Heavy Search: Prioritize deep logical flaws that traditional static analysis tools often miss, such as flawed economic incentives, incorrect state updates, or integration-specific vulnerabilities (e.g., ERC-20 token standard edge cases like fee-on-transfer or rebasing tokens).

**Pruning:** Only return findings that represent a genuine, realistically exploitable risk to funds, contract availability, or administrative integrity.

**Code:**

\hspace*{1em}\texttt{\detokenize{[CONTRACT_CODE]}}
\end{quote}

\subsection{CoT + Slither Prompt}

Reasoning-oriented prompt with Slither grounding.

\begin{quote}
\small\raggedright
**Role:** You are a senior Smart Contract Security Auditor specializing in Solidity and EVM-level exploits.

**Task:** Conduct a high-precision security analysis of the provided contract(s).

**Execution Directives:**

1. Baseline Triage: I have provided the Slither static analysis output below. Use this to identify immediate "hotspots" in the code. In your internal reasoning, evaluate if these detections are true positives or if the contract's specific business logic renders them non-exploitable.
2. Independent Invariant Mapping: Disregard the Slither output for a moment and independently identify the critical security invariants of this contract (e.g., "The contract must never allow withdrawals that exceed the msg.sender's accounted balance").
3. The "Blind Spot" Search: Proactively search for high-level logical vulnerabilities that Slither is fundamentally unable to detect, such as flawed economic math, incorrect state updates across multiple transactions, or subtle integration risks with external protocols.
4. Adversarial Pressure Test: Attempt to "disprove" the Slither findings. If Slither flags a re-entrancy vulnerability, can you prove it is actually benign? Conversely, if Slither flags nothing in a critical function like withdraw(), can you find a sequence of transactions that breaks it anyway?

**Pruning:** Only return findings that represent a genuine, realistically exploitable risk to funds, contract availability, or administrative integrity.

**Code:**

\hspace*{1em}\texttt{\detokenize{[CONTRACT_CODE]}}

**Slither Output:**

\hspace*{1em}\texttt{\detokenize{[SLITHER_DETECTIONS]}}
\end{quote}

\subsection{Structured Output Prompt}

System prompt used to enforce JSON-conformant structured output when native schema enforcement is unavailable.

\begin{quote}
\small\raggedright
The user will provide input text. Please analyze the content and map the relevant information into the structured JSON format provided below.

**JSON Schema:**

\hspace*{1em}\texttt{\detokenize{[JSON_SCHEMA]}}

**Instructions:**
1. The output must strictly conform to the schema provided.
2. Return ONLY the raw JSON.
3. Do not wrap the response in markdown backticks or provide any explanations.
4. If no relevant data is found for a field, use 'null' or an empty array.
\end{quote}

\section{Vulnerability Subcategories}
\label{app:defect-subcategories}

The filtered benchmark retains the following defect subcategories, grouped by the high-level taxonomy in Table~\ref{tab:defect_categories}.

\subsection*{1 Malicious Env., Transactions or Input}
\begin{itemize}[leftmargin=*,nosep]
    \item \defectCategory{1A Reentrancy}
    \item \defectCategory{1B Call To The Unknown}
    \item \defectCategory{1C Exact Balance Dependency}
    \item \defectCategory{1D Improper Data Validation}
    \item \defectCategory{1E Vulnerable DELEGATECALL}
\end{itemize}

\subsection*{2 Blockchain/Environment Dependency}
\begin{itemize}[leftmargin=*,nosep]
    \item \defectCategory{2A Timestamp Dependency}
    \item \defectCategory{2B Transaction-Ordering Dependency (TOD)}
    \item \defectCategory{2C Bad Random Number Generation}
    \item \defectCategory{2D Improper Private Variable Secret}
    \item \defectCategory{2E Unpredictable State (Dynamic Libraries)}
    \item \defectCategory{2F Block Hash Dependency}
    \item \defectCategory{2G Block Number Dependency}
\end{itemize}

\subsection*{3 Exception \& Error Handling Disorders}
\begin{itemize}[leftmargin=*,nosep]
    \item \defectCategory{3A Unchecked Low Level Call or Send Return Values}
    \item \begingroup\settowidth{\dimen0}{\defectCategory{3B Unexpected Throw and Revert}}%
    \makebox[0pt][l]{\defectCategory{3B Unexpected Throw and Revert}}%
    \rule[0.45ex]{\dimen0}{0.4pt}%
    \endgroup\hspace{0.4em}\textit{(excluded due to low classifier accuracy)}
    \item \defectCategory{3C Mishandled Out-Of-Gas Exception}
    \item \begingroup\settowidth{\dimen0}{\defectCategory{3D Assert, Require or Revert Violation}}%
    \makebox[0pt][l]{\defectCategory{3D Assert, Require or Revert Violation}}%
    \rule[0.45ex]{\dimen0}{0.4pt}%
    \endgroup\hspace{0.4em}\textit{(excluded due to low classifier accuracy)}
\end{itemize}

\subsection*{4 Denial of Service}
\begin{itemize}[leftmargin=*,nosep]
    \item \defectCategory{4A Frozen Ether}
    \item \defectCategory{4B Ether Lost In Transfer}
    \item \defectCategory{4C DoS With Block Gas Limit Reached}
    \item \defectCategory{4D DoS By Exception Inside Loop}
    \item \defectCategory{4E Insufficient Gas Griefing}
\end{itemize}

\subsection*{5 Resource Consumption \& Gas Issues}
\begin{itemize}[leftmargin=*,nosep]
    \item \defectCategory{5A Gas Costly Loops}
    \item \defectCategory{5B Gas Costly Pattern}
    \item \defectCategory{5C High Gas Consumption Of Variable Datatype Or Declaration}
    \item \defectCategory{5D High Gas Consumption Function Type}
    \item \defectCategory{5E Under-Priced Opcodes}
\end{itemize}

\subsection*{6 Auth. \& Access Control Vulnerabilities}
\begin{itemize}[leftmargin=*,nosep]
    \item \defectCategory{6A Authorization Via Transaction Origin}
    \item \defectCategory{6B Unauthorized Accessibility Due To Wrong Function Or State Variable Visibility}
    \item \defectCategory{6C Unprotected Self-Destruction}
    \item \defectCategory{6D Unauthorized Ether Withdrawal}
    \item \defectCategory{6E Signature-Based Vulnerabilities}
\end{itemize}

\subsection*{7 Arithmetic Bugs}
\begin{itemize}[leftmargin=*,nosep]
    \item \defectCategory{7A Integer Over Or Underflow}
    \item \defectCategory{7B No Division by Zero check}
    \item \defectCategory{7C Integer Division Remainder}
\end{itemize}

\subsection*{8 Bad Coding and Language Specifics}
\begin{itemize}[leftmargin=*,nosep]
    \item \defectCategory{8A Type Cast}
    \item \defectCategory{8B Coding Error}
    \item \defectCategory{8C Function Ordering}
    \item \defectCategory{8D Deprecated Source Language Features}
    \item \defectCategory{8F Use Of Assembly}
    \item \defectCategory{8G Incorrect Inheritance Order}
    \item \defectCategory{8H Variable Shadowing}
    \item \defectCategory{8I Single Quotes String}
    \item \defectCategory{8J Dead Code}
    \item \defectCategory{8K Violation Of Checks-Effects-Interaction Pattern}
    \item \defectCategory{8L Inadequate Or Incorrect Logging Or Documentation}
\end{itemize}

\subsection*{9 Environment Configuration Issues}
\begin{itemize}[leftmargin=*,nosep]
    \item \defectCategory{9A Missing Short Address Or Zero Address validation}
    \item \defectCategory{9B Outdated Compiler Version}
    \item \defectCategory{9C Floating Or No Pragma}
    \item \defectCategory{9D Token API Violation}
    \item \defectCategory{9F Hardcoded Gas}
\end{itemize}

\subsection*{10 Eliminated/Deprecated Vulnerabilities}
\begin{itemize}[leftmargin=*,nosep]
    \item \begingroup\settowidth{\dimen0}{\defectCategory{10A Callstack Depth Limit}}%
    \makebox[0pt][l]{\defectCategory{10A Callstack Depth Limit}}%
    \rule[0.45ex]{\dimen0}{0.4pt}%
    \endgroup\hspace{0.4em}\textit{(excluded due to low classifier accuracy)}
    \item \defectCategory{10C Erroneous Constructor Name}
\end{itemize}
\section{Allowed Close Matches for Classification}
\label{app:close-match-pairs}

The following close-match pairs were accepted during automatic classification.

\begin{table}[H]
\caption{Close-match pairs accepted during automatic classification}
\label{tab:close-match-pairs}
\centering
\normalsize
\begin{tabular}{|l|l|}
\hline
\textbf{Exact Match} & \textbf{Alternative} \\
\hline
\parbox[t]{0.42\columnwidth}{\raggedright\defectCategory{Call To The Unknown}} & \parbox[t]{0.42\columnwidth}{\raggedright\defectCategory{Vulnerable DELEGATECALL}} \\
\hline
\parbox[t]{0.42\columnwidth}{\raggedright\defectCategory{Gas Costly Loops}} & \parbox[t]{0.42\columnwidth}{\raggedright\defectCategory{Gas Costly Pattern}} \\
\hline
\parbox[t]{0.42\columnwidth}{\raggedright\defectCategory{Gas Costly Pattern}} & \parbox[t]{0.42\columnwidth}{\raggedright\defectCategory{Gas Costly Loops}} \\
\hline
\parbox[t]{0.42\columnwidth}{\raggedright\defectCategory{High Gas Consumption Function Type}} & \parbox[t]{0.42\columnwidth}{\raggedright\defectCategory{High Gas Consumption Variable}} \\
\hline
\parbox[t]{0.42\columnwidth}{\raggedright\defectCategory{High Gas Consumption Variable}} & \parbox[t]{0.42\columnwidth}{\raggedright\defectCategory{High Gas Consumption Function Type}} \\
\hline
\parbox[t]{0.42\columnwidth}{\raggedright\defectCategory{Violation Of Checks-Effects-Interaction Pattern}} & \parbox[t]{0.42\columnwidth}{\raggedright\defectCategory{Reentrancy}} \\
\hline
\parbox[t]{0.42\columnwidth}{\raggedright\defectCategory{Vulnerable DELEGATECALL}} & \parbox[t]{0.42\columnwidth}{\raggedright\defectCategory{Call To The Unknown}} \\
\hline
\end{tabular}
\end{table}
\section{API Providers Used for Model Access}
\label{app:api-providers}

The following table lists the API providers used in the study and the models accessed through each provider.

\begin{table}[H]
\caption{API providers and models included in our study}
\label{tab:api-providers}
\centering
\normalsize
\begin{tabular}{|l|l|}
\hline
\textbf{API Provider} & \textbf{Models} \\
\hline
\parbox[t]{0.24\columnwidth}{\raggedright DeepSeek} & \parbox[t]{0.62\columnwidth}{\raggedright DeepSeek Chat, DeepSeek Reasoner, DeepSeek R1 14b} \\
\hline
\parbox[t]{0.24\columnwidth}{\raggedright Mistral} & \parbox[t]{0.62\columnwidth}{\raggedright Magistral Medium, Magistral Small (API), Devstral, Devstral Small (API)} \\
\hline
\parbox[t]{0.24\columnwidth}{\raggedright xAI} & \parbox[t]{0.62\columnwidth}{\raggedright Grok 4.1 Fast Reasoning, Grok 4.1 Fast Non-Reasoning} \\
\hline
\parbox[t]{0.24\columnwidth}{\raggedright OpenRouter} & \parbox[t]{0.62\columnwidth}{\raggedright GPT Oss 120b, GPT Oss 20b (API)} \\
\hline
\parbox[t]{0.24\columnwidth}{\raggedright Alibaba Cloud} & \parbox[t]{0.62\columnwidth}{\raggedright Qwen3 30b (thinking), Qwen3 30b (instruct), Qwen3 235b (thinking), Qwen3 235b (instruct)} \\
\hline
\parbox[t]{0.24\columnwidth}{\raggedright Ollama} & \parbox[t]{0.62\columnwidth}{\raggedright DeepSeek R1 14b (local), Devstral Small (local), GPT Oss 20b (local), Magistral 24b (local)} \\
\hline
\end{tabular}
\end{table}
\section{Experimental Results}
\label{app:experiment-results}
This appendix contains the detailed results for the five experiments. Each experiment is organized into three tables: detection metrics, confusion-matrix counts, and reliability.

The first table for each experiment reports detection metrics.
\begin{description}[leftmargin=*,nosep,style=nextline]
	\item[Model] identifies the evaluated LLM.
	\item[Recall] gives the proportion of positive instances correctly identified by the model.
	\item[False Positive Rate] gives the proportion of negative instances incorrectly classified as positive.
	\item[Balanced Accuracy] gives the average of recall and the true-negative rate.
	\item[Precision] gives the proportion of positive predictions that are correct.
\end{description}

The second table reports the confusion-matrix counts used to compute the metrics.
\begin{description}[leftmargin=*,nosep,style=nextline]
	\item[Model] identifies the evaluated LLM.
	\item[True Positives] counts positive instances correctly identified as positive.
	\item[False Positives] counts negative instances incorrectly identified as positive.
	\item[True Negatives] counts negative instances correctly identified as negative.
	\item[False Negatives] counts positive instances incorrectly identified as negative.
\end{description}

The third table summarizes reliability and execution behavior.
\begin{description}[leftmargin=*,nosep,style=nextline]
	\item[Model] identifies the evaluated LLM.
	\item[Timeouts] counts runs that exceeded the allowed time limit.
	\item[Invalid Results] counts instances in which the model did not return output conforming to the expected structure.
	\item[Detections] counts the total number of detections produced by the model.
	\item[Average runtime in milliseconds] gives the mean response time per contract.
	\item[Errors] counts non-transient errors, that is, failures that persisted across retries. In our experiments, transient errors were typically provider-side overloads, whereas non-transient errors were observed mainly in locally deployed models because of hardware limitations on the MacBook M5 system with 24 GB unified memory.
\end{description}

\begin{table}[H]
\centering
\caption{Experiment 1 detection metrics}
\label{tab:exp1-abbreviated-metrics}
\resizebox{\columnwidth}{!}{%
\begin{tabular}{|l|c|c|c|c|}
\hline
\textbf{Model} & \textbf{Recall} & \textbf{False Positive Rate} & \textbf{Balanced Accuracy} & \textbf{Precision} \\
\hline
GPT Oss 120b & 0.35 & 0.35 & 0.5 & 0.5 \\
\hline
GPT Oss 20b API & 0.21 & 0.12 & 0.54 & 0.63 \\
\hline
GPT Oss 20b Local & 0.36 & 0.14 & 0.6 & 0.71 \\
\hline
Grok 4.1 fast non-reasoning & 0.67 & 0.3 & 0.68 & 0.68 \\
\hline
Grok 4.1 fast reasoning & 0.68 & 0.23 & 0.72 & 0.74 \\
\hline
Devstral & 0.57 & 0.4 & 0.58 & 0.58 \\
\hline
Magistral Medium & 0.52 & 0.28 & 0.61 & 0.64 \\
\hline
Magistral 24b API & 0.4 & 0.25 & 0.57 & 0.62 \\
\hline
Devstral Small API & 0.55 & 0.38 & 0.58 & 0.59 \\
\hline
Magistral 24b Local & 0.52 & 0.34 & 0.59 & 0.6 \\
\hline
Devstral Small Local & 0.48 & 0.32 & 0.57 & 0.59 \\
\hline
Qwen3 235b thinking & 0.55 & 0.17 & 0.69 & 0.76 \\
\hline
Qwen3 30b thinking API & 0.43 & 0.14 & 0.64 & 0.74 \\
\hline
Qwen3 235b instruct & 0.62 & 0.37 & 0.62 & 0.62 \\
\hline
Qwen3 30b instruct API & 0.6 & 0.51 & 0.54 & 0.54 \\
\hline
DeepSeek Chat & 0.69 & 0.38 & 0.65 & 0.64 \\
\hline
DeepSeek Reasoner & 0.64 & 0.21 & 0.71 & 0.75 \\
\hline
DeepSeek R1 14b Local & 0.46 & 0.21 & 0.62 & 0.68 \\
\hline
\end{tabular}%
}
\end{table}

\begin{table}[H]
\centering
\caption{Experiment 1 TP, FP, TN, and FN counts}
\label{tab:exp1-counts}
\resizebox{\columnwidth}{!}{%
\begin{tabular}{|l|c|c|c|c|}
\hline
\textbf{Model} & \textbf{True Positives} & \textbf{False Positives} & \textbf{True Negatives} & \textbf{False Negatives} \\
\hline
GPT Oss 120b & 31 & 31 & 57 & 57 \\
\hline
GPT Oss 20b API & 19 & 11 & 77 & 69 \\
\hline
GPT Oss 20b Local & 32 & 13 & 75 & 56 \\
\hline
Grok 4.1 fast non-reasoning & 59 & 27 & 61 & 29 \\
\hline
Grok 4.1 fast reasoning & 60 & 21 & 67 & 28 \\
\hline
Devstral & 51 & 36 & 52 & 37 \\
\hline
Magistral Medium & 46 & 25 & 63 & 42 \\
\hline
Magistral 24b API & 36 & 22 & 66 & 52 \\
\hline
Devstral Small API & 49 & 34 & 54 & 39 \\
\hline
Magistral 24b Local & 46 & 30 & 58 & 42 \\
\hline
Devstral Small Local & 43 & 29 & 59 & 45 \\
\hline
Qwen3 235b thinking & 49 & 15 & 73 & 39 \\
\hline
Qwen3 30b thinking API & 38 & 13 & 75 & 50 \\
\hline
Qwen3 235b instruct & 55 & 33 & 55 & 33 \\
\hline
Qwen3 30b instruct API & 53 & 45 & 43 & 35 \\
\hline
DeepSeek Chat & 61 & 34 & 54 & 27 \\
\hline
DeepSeek Reasoner & 57 & 19 & 69 & 31 \\
\hline
DeepSeek R1 14b Local & 41 & 19 & 69 & 47 \\
\hline
\end{tabular}%
}
\end{table}

\begin{table}[H]
\centering
\caption{Experiment 1 reliability}
\label{tab:exp1-reliability}
\resizebox{\columnwidth}{!}{%
\begin{tabular}{|l|c|c|c|c|c|}
\hline
\textbf{Model} & \textbf{Timeouts} & \textbf{Invalid Results} & \textbf{Detections} & \textbf{Average runtime (ms)} & \textbf{Errors} \\
\hline
GPT Oss 120b & 0 & 47 & 811 & 26246.45 & 0 \\
\hline
GPT Oss 20b API & 0 & 24 & 297 & 19687.48 & 0 \\
\hline
GPT Oss 20b Local & 15 & 19 & 335 & 82885.19 & 2 \\
\hline
Grok 4.1 fast non-reasoning & 0 & 0 & 803 & 4527.81 & 0 \\
\hline
Grok 4.1 fast reasoning & 0 & 0 & 707 & 21461.32 & 0 \\
\hline
Devstral & 0 & 0 & 728 & 7751.04 & 0 \\
\hline
Magistral Medium & 0 & 22 & 422 & 10743.78 & 0 \\
\hline
Magistral 24b API & 0 & 13 & 430 & 2895.19 & 0 \\
\hline
Devstral Small API & 0 & 5 & 593 & 2489.17 & 0 \\
\hline
Magistral 24b Local & 0 & 9 & 432 & 39808.92 & 0 \\
\hline
Devstral Small Local & 0 & 11 & 587 & 58282 & 11 \\
\hline
Qwen3 235b thinking & 0 & 0 & 389 & 72305.21 & 0 \\
\hline
Qwen3 30b thinking API & 0 & 2 & 211 & 21818.17 & 0 \\
\hline
Qwen3 235b instruct & 0 & 5 & 534 & 6913.48 & 0 \\
\hline
Qwen3 30b instruct API & 0 & 7 & 420 & 5116.01 & 1 \\
\hline
DeepSeek Chat & 0 & 0 & 680 & 16232.74 & 0 \\
\hline
DeepSeek Reasoner & 0 & 0 & 478 & 77573.37 & 0 \\
\hline
DeepSeek R1 14b Local & 0 & 15 & 233 & 62654.03 & 0 \\
\hline
\end{tabular}%
}
\end{table}

\begin{table}[H]
\centering
\caption{Experiment 2 detection metrics}
\label{tab:exp2-abbreviated-metrics}
\resizebox{\columnwidth}{!}{%
\begin{tabular}{|l|c|c|c|c|}
\hline
\textbf{Model} & \textbf{Recall} & \textbf{False Positive Rate} & \textbf{Balanced Accuracy} & \textbf{Precision} \\
\hline
GPT Oss 120b & 0.26 & 0.19 & 0.53 & 0.57 \\
\hline
GPT Oss 20b API & 0.15 & 0.09 & 0.53 & 0.63 \\
\hline
GPT Oss 20b Local & 0.23 & 0.11 & 0.56 & 0.67 \\
\hline
Grok 4.1 fast non-reasoning & 0.43 & 0.31 & 0.55 & 0.57 \\
\hline
Grok 4.1 fast reasoning & 0.43 & 0.27 & 0.57 & 0.61 \\
\hline
Devstral & 0.38 & 0.29 & 0.54 & 0.56 \\
\hline
Magistral Medium & 0.28 & 0.23 & 0.52 & 0.54 \\
\hline
Magistral 24b API & 0.23 & 0.19 & 0.52 & 0.55 \\
\hline
Devstral Small API & 0.32 & 0.22 & 0.55 & 0.59 \\
\hline
Magistral 24b Local & 0.29 & 0.22 & 0.53 & 0.56 \\
\hline
Devstral Small Local & 0.3 & 0.25 & 0.52 & 0.55 \\
\hline
Qwen3 235b thinking & 0.31 & 0.22 & 0.54 & 0.58 \\
\hline
Qwen3 30b thinking API & 0.19 & 0.1 & 0.54 & 0.65 \\
\hline
Qwen3 235b instruct & 0.39 & 0.34 & 0.52 & 0.53 \\
\hline
Qwen3 30b instruct API & 0.28 & 0.29 & 0.49 & 0.49 \\
\hline
DeepSeek R1 14b Local & 0.22 & 0.19 & 0.51 & 0.54 \\
\hline
DeepSeek Chat & 0.38 & 0.35 & 0.51 & 0.52 \\
\hline
DeepSeek Reasoner & 0.37 & 0.21 & 0.57 & 0.63 \\
\hline
\end{tabular}%
}
\end{table}

\begin{table}[H]
\centering
\caption{Experiment 2 TP, FP, TN, and FN counts}
\label{tab:exp2-counts}
\resizebox{\columnwidth}{!}{%
\begin{tabular}{|l|c|c|c|c|}
\hline
\textbf{Model} & \textbf{True Positives} & \textbf{False Positives} & \textbf{True Negatives} & \textbf{False Negatives} \\
\hline
GPT Oss 120b & 23 & 17 & 71 & 65 \\
\hline
GPT Oss 20b API & 14 & 8 & 80 & 74 \\
\hline
GPT Oss 20b Local & 21 & 10 & 78 & 67 \\
\hline
Grok 4.1 fast non-reasoning & 38 & 28 & 60 & 50 \\
\hline
Grok 4.1 fast reasoning & 38 & 24 & 64 & 50 \\
\hline
Devstral & 34 & 26 & 62 & 54 \\
\hline
Magistral Medium & 25 & 21 & 67 & 63 \\
\hline
Magistral 24b API & 21 & 17 & 71 & 67 \\
\hline
Devstral Small API & 29 & 20 & 68 & 59 \\
\hline
Magistral 24b Local & 26 & 20 & 68 & 62 \\
\hline
Devstral Small Local & 27 & 22 & 66 & 61 \\
\hline
Qwen3 235b thinking & 28 & 20 & 68 & 60 \\
\hline
Qwen3 30b thinking API & 17 & 9 & 79 & 71 \\
\hline
Qwen3 235b instruct & 35 & 30 & 58 & 53 \\
\hline
Qwen3 30b instruct API & 25 & 26 & 62 & 63 \\
\hline
DeepSeek R1 14b Local & 20 & 17 & 71 & 68 \\
\hline
DeepSeek Chat & 34 & 31 & 57 & 54 \\
\hline
DeepSeek Reasoner & 33 & 19 & 69 & 55 \\
\hline
\end{tabular}%
}
\end{table}

\begin{table}[H]
\centering
\caption{Experiment 2 reliability}
\label{tab:exp2-reliability}
\resizebox{\columnwidth}{!}{%
\begin{tabular}{|l|c|c|c|c|c|}
\hline
\textbf{Model} & \textbf{Timeouts} & \textbf{Invalid Results} & \textbf{Detections} & \textbf{Average runtime (ms)} & \textbf{Errors} \\
\hline
GPT Oss 120b & 0 & 53 & 821 & 30509.31 & 0 \\
\hline
GPT Oss 20b API & 0 & 38 & 385 & 38906.93 & 0 \\
\hline
GPT Oss 20b Local & 14 & 26 & 367 & 84266.04 & 1 \\
\hline
Grok 4.1 fast non-reasoning & 0 & 0 & 871 & 6571.7 & 0 \\
\hline
Grok 4.1 fast reasoning & 0 & 1 & 822 & 32698.62 & 0 \\
\hline
Devstral & 0 & 0 & 908 & 8258.89 & 0 \\
\hline
Magistral Medium & 0 & 25 & 573 & 8050.61 & 0 \\
\hline
Magistral 24b API & 0 & 17 & 486 & 4737.48 & 0 \\
\hline
Devstral Small API & 0 & 12 & 669 & 3084.46 & 0 \\
\hline
Magistral 24b Local & 0 & 5 & 566 & 47013.09 & 0 \\
\hline
Devstral Small Local & 0 & 10 & 687 & 65407.06 & 10 \\
\hline
Qwen3 235b thinking & 0 & 0 & 458 & 82007.36 & 0 \\
\hline
Qwen3 30b thinking API & 0 & 4 & 253 & 25308.17 & 0 \\
\hline
Qwen3 235b instruct & 0 & 8 & 702 & 8536.02 & 0 \\
\hline
Qwen3 30b instruct API & 0 & 11 & 602 & 7375.38 & 0 \\
\hline
DeepSeek R1 14b Local & 0 & 4 & 311 & 67691.12 & 0 \\
\hline
DeepSeek Chat & 0 & 0 & 831 & 25226.73 & 0 \\
\hline
DeepSeek Reasoner & 0 & 0 & 585 & 86275.03 & 0 \\
\hline
\end{tabular}%
}
\end{table}

\begin{table}[H]
\centering
\caption{Experiment 3 detection metrics}
\label{tab:exp3-abbreviated-metrics}
\resizebox{\columnwidth}{!}{%
\begin{tabular}{|l|c|c|c|c|}
\hline
\textbf{Model} & \textbf{Recall} & \textbf{False Positive Rate} & \textbf{Balanced Accuracy} & \textbf{Precision} \\
\hline
GPT Oss 120b & 0.07 & 0.02 & 0.52 & 0.77 \\
\hline
GPT Oss 20b API & 0.11 & 0.12 & 0.49 & 0.47 \\
\hline
GPT Oss 20b Local & 0.18 & 0.02 & 0.57 & 0.88 \\
\hline
Grok 4.1 fast non-reasoning & 0.39 & 0.26 & 0.56 & 0.6 \\
\hline
Grok 4.1 fast reasoning & 0.25 & 0.11 & 0.56 & 0.68 \\
\hline
Devstral & 0.27 & 0.23 & 0.51 & 0.53 \\
\hline
Magistral Medium & 0.25 & 0.23 & 0.5 & 0.51 \\
\hline
Magistral 24b API & 0.15 & 0.13 & 0.51 & 0.53 \\
\hline
Devstral Small API & 0.23 & 0.2 & 0.51 & 0.53 \\
\hline
Magistral 24b Local & 0.18 & 0.18 & 0.5 & 0.5 \\
\hline
Devstral Small Local & 0.19 & 0.19 & 0.5 & 0.5 \\
\hline
Qwen3 235b thinking & 0.23 & 0.09 & 0.57 & 0.72 \\
\hline
Qwen3 30b thinking API & 0.11 & 0.06 & 0.52 & 0.62 \\
\hline
Qwen3 235b instruct & 0.34 & 0.28 & 0.52 & 0.54 \\
\hline
Qwen3 30b instruct API & 0.22 & 0.18 & 0.52 & 0.55 \\
\hline
DeepSeek Chat & 0.37 & 0.29 & 0.53 & 0.55 \\
\hline
DeepSeek Reasoner & 0.29 & 0.15 & 0.56 & 0.65 \\
\hline
DeepSeek R1 14b Local & 0.17 & 0.14 & 0.51 & 0.53 \\
\hline
\end{tabular}%
}
\end{table}

\begin{table}[H]
\centering
\caption{Experiment 3 TP, FP, TN, and FN counts}
\label{tab:exp3-counts}
\resizebox{\columnwidth}{!}{%
\begin{tabular}{|l|c|c|c|c|}
\hline
\textbf{Model} & \textbf{True Positives} & \textbf{False Positives} & \textbf{True Negatives} & \textbf{False Negatives} \\
\hline
GPT Oss 120b & 7 & 2 & 86 & 81 \\
\hline
GPT Oss 20b API & 10 & 11 & 77 & 78 \\
\hline
GPT Oss 20b Local & 16 & 2 & 86 & 72 \\
\hline
Grok 4.1 fast non-reasoning & 35 & 23 & 65 & 53 \\
\hline
Grok 4.1 fast reasoning & 22 & 10 & 78 & 66 \\
\hline
Devstral & 24 & 21 & 67 & 64 \\
\hline
Magistral Medium & 22 & 21 & 67 & 66 \\
\hline
Magistral 24b API & 14 & 12 & 76 & 74 \\
\hline
Devstral Small API & 21 & 18 & 70 & 67 \\
\hline
Magistral 24b Local & 16 & 16 & 72 & 72 \\
\hline
Devstral Small Local & 17 & 17 & 71 & 71 \\
\hline
Qwen3 235b thinking & 21 & 8 & 80 & 67 \\
\hline
Qwen3 30b thinking API & 10 & 6 & 82 & 78 \\
\hline
Qwen3 235b instruct & 30 & 25 & 63 & 58 \\
\hline
Qwen3 30b instruct API & 20 & 16 & 72 & 68 \\
\hline
DeepSeek Chat & 33 & 26 & 62 & 55 \\
\hline
DeepSeek Reasoner & 26 & 14 & 74 & 62 \\
\hline
DeepSeek R1 14b Local & 15 & 13 & 75 & 73 \\
\hline
\end{tabular}%
}
\end{table}

\begin{table}[H]
\centering
\caption{Experiment 3 reliability}
\label{tab:exp3-reliability}
\resizebox{\columnwidth}{!}{%
\begin{tabular}{|l|c|c|c|c|c|}
\hline
\textbf{Model} & \textbf{Timeouts} & \textbf{Invalid Results} & \textbf{Detections} & \textbf{Average runtime (ms)} & \textbf{Errors} \\
\hline
GPT Oss 120b & 0 & 136 & 100 & 33082.94 & 0 \\
\hline
GPT Oss 20b API & 0 & 19 & 283 & 20209.19 & 0 \\
\hline
GPT Oss 20b Local & 24 & 20 & 130 & 92362.95 & 1 \\
\hline
Grok 4.1 fast non-reasoning & 0 & 0 & 616 & 5859.35 & 0 \\
\hline
Grok 4.1 fast reasoning & 0 & 0 & 367 & 30982.92 & 0 \\
\hline
Devstral & 0 & 0 & 719 & 19870.96 & 0 \\
\hline
Magistral Medium & 0 & 25 & 454 & 9417.3 & 0 \\
\hline
Magistral 24b API & 0 & 32 & 294 & 2949.55 & 0 \\
\hline
Devstral Small API & 0 & 6 & 562 & 2471.74 & 0 \\
\hline
Magistral 24b Local & 0 & 15 & 456 & 44624.74 & 0 \\
\hline
Devstral Small Local & 0 & 15 & 547 & 58759.19 & 15 \\
\hline
Qwen3 235b thinking & 0 & 0 & 267 & 89767.63 & 0 \\
\hline
Qwen3 30b thinking API & 0 & 0 & 133 & 27868.15 & 0 \\
\hline
Qwen3 235b instruct & 0 & 2 & 570 & 8527.46 & 0 \\
\hline
Qwen3 30b instruct API & 0 & 9 & 448 & 6838.55 & 0 \\
\hline
DeepSeek Chat & 0 & 0 & 699 & 20865.5 & 0 \\
\hline
DeepSeek Reasoner & 0 & 0 & 436 & 90862.1 & 0 \\
\hline
DeepSeek R1 14b Local & 3 & 5 & 268 & 78296.52 & 0 \\
\hline
\end{tabular}%
}
\end{table}

\begin{table}[H]
\centering
\caption{Experiment 4 detection metrics}
\label{tab:exp4-abbreviated-metrics}
\resizebox{\columnwidth}{!}{%
\begin{tabular}{|l|c|c|c|c|}
\hline
\textbf{Model} & \textbf{Recall} & \textbf{False Positive Rate} & \textbf{Balanced Accuracy} & \textbf{Precision} \\
\hline
GPT Oss 120b & 0.34 & 0.11 & 0.61 & 0.75 \\
\hline
GPT Oss 20b API & 0.22 & 0.07 & 0.57 & 0.74 \\
\hline
GPT Oss 20b Local & 0.3 & 0.14 & 0.57 & 0.67 \\
\hline
Grok 4.1 fast non-reasoning & 0.46 & 0.35 & 0.55 & 0.56 \\
\hline
Grok 4.1 fast reasoning & 0.45 & 0.27 & 0.59 & 0.62 \\
\hline
Devstral & 0.47 & 0.31 & 0.57 & 0.6 \\
\hline
Magistral Medium & 0.36 & 0.3 & 0.52 & 0.54 \\
\hline
Magistral 24b API & 0.3 & 0.23 & 0.53 & 0.56 \\
\hline
Devstral Small API & 0.35 & 0.27 & 0.53 & 0.56 \\
\hline
Magistral 24b Local & 0.32 & 0.22 & 0.55 & 0.59 \\
\hline
Devstral Small Local & 0.4 & 0.28 & 0.56 & 0.59 \\
\hline
Qwen3 235b thinking & 0.38 & 0.22 & 0.57 & 0.62 \\
\hline
Qwen3 30b thinking API & 0.34 & 0.11 & 0.61 & 0.75 \\
\hline
Qwen3 235b instruct & 0.43 & 0.29 & 0.56 & 0.59 \\
\hline
Qwen3 30b instruct API & 0.31 & 0.25 & 0.53 & 0.56 \\
\hline
DeepSeek Chat & 0.51 & 0.4 & 0.55 & 0.55 \\
\hline
DeepSeek Reasoner & 0.46 & 0.22 & 0.61 & 0.67 \\
\hline
DeepSeek R1 14b Local & 0.3 & 0.12 & 0.59 & 0.71 \\
\hline
\end{tabular}%
}
\end{table}

\begin{table}[H]
\centering
\caption{Experiment 4 TP, FP, TN, and FN counts}
\label{tab:exp4-counts}
\resizebox{\columnwidth}{!}{%
\begin{tabular}{|l|c|c|c|c|}
\hline
\textbf{Model} & \textbf{True Positives} & \textbf{False Positives} & \textbf{True Negatives} & \textbf{False Negatives} \\
\hline
GPT Oss 120b & 30 & 10 & 78 & 58 \\
\hline
GPT Oss 20b API & 20 & 7 & 81 & 68 \\
\hline
GPT Oss 20b Local & 27 & 13 & 75 & 61 \\
\hline
Grok 4.1 fast non-reasoning & 41 & 31 & 57 & 47 \\
\hline
Grok 4.1 fast reasoning & 40 & 24 & 64 & 48 \\
\hline
Devstral & 42 & 28 & 60 & 46 \\
\hline
Magistral Medium & 32 & 27 & 61 & 56 \\
\hline
Magistral 24b API & 27 & 21 & 67 & 61 \\
\hline
Devstral Small API & 31 & 24 & 64 & 57 \\
\hline
Magistral 24b Local & 29 & 20 & 68 & 59 \\
\hline
Devstral Small Local & 36 & 25 & 63 & 52 \\
\hline
Qwen3 235b thinking & 34 & 20 & 68 & 54 \\
\hline
Qwen3 30b thinking API & 30 & 10 & 78 & 58 \\
\hline
Qwen3 235b instruct & 38 & 26 & 62 & 50 \\
\hline
Qwen3 30b instruct API & 28 & 22 & 66 & 60 \\
\hline
DeepSeek Chat & 45 & 36 & 52 & 43 \\
\hline
DeepSeek Reasoner & 41 & 20 & 68 & 47 \\
\hline
DeepSeek R1 14b Local & 27 & 11 & 77 & 61 \\
\hline
\end{tabular}%
}
\end{table}

\begin{table}[H]
\centering
\caption{Experiment 4 reliability}
\label{tab:exp4-reliability}
\resizebox{\columnwidth}{!}{%
\begin{tabular}{|l|c|c|c|c|c|}
\hline
\textbf{Model} & \textbf{Timeouts} & \textbf{Invalid Results} & \textbf{Detections} & \textbf{Average runtime (ms)} & \textbf{Errors} \\
\hline
GPT Oss 120b & 0 & 19 & 476 & 22133.51 & 0 \\
\hline
GPT Oss 20b API & 0 & 9 & 239 & 17271.72 & 0 \\
\hline
GPT Oss 20b Local & 19 & 31 & 549 & 122301.42 & 3 \\
\hline
Grok 4.1 fast non-reasoning & 0 & 0 & 1212 & 5824.03 & 0 \\
\hline
Grok 4.1 fast reasoning & 0 & 1 & 930 & 24920.38 & 0 \\
\hline
Devstral & 0 & 0 & 1059 & 15997.43 & 0 \\
\hline
Magistral Medium & 0 & 14 & 641 & 14881.31 & 0 \\
\hline
Magistral 24b API & 0 & 18 & 680 & 5328.39 & 0 \\
\hline
Devstral Small API & 0 & 8 & 812 & 3155.15 & 0 \\
\hline
Magistral 24b Local & 0 & 12 & 653 & 55521.33 & 0 \\
\hline
Devstral Small Local & 0 & 3 & 856 & 78918.41 & 3 \\
\hline
Qwen3 235b thinking & 0 & 0 & 601 & 91463.6 & 0 \\
\hline
Qwen3 30b thinking API & 0 & 3 & 397 & 25569.06 & 0 \\
\hline
Qwen3 235b instruct & 0 & 2 & 783 & 9638.89 & 0 \\
\hline
Qwen3 30b instruct API & 0 & 18 & 821 & 11242.03 & 0 \\
\hline
DeepSeek Chat & 0 & 0 & 1015 & 32165.43 & 0 \\
\hline
DeepSeek Reasoner & 1 & 0 & 751 & 94758.32 & 0 \\
\hline
DeepSeek R1 14b Local & 0 & 4 & 384 & 70865.82 & 0 \\
\hline
\end{tabular}%
}
\end{table}

\begin{table}[H]
\centering
\caption{Experiment 5 detection metrics}
\label{tab:exp5-abbreviated-metrics}
\resizebox{\columnwidth}{!}{%
\begin{tabular}{|l|c|c|c|c|}
\hline
\textbf{Model} & \textbf{Recall} & \textbf{False Positive Rate} & \textbf{Balanced Accuracy} & \textbf{Precision} \\
\hline
GPT Oss 120b & 0.21 & 0.1 & 0.55 & 0.67 \\
\hline
GPT Oss 20b API & 0.2 & 0.07 & 0.56 & 0.72 \\
\hline
GPT Oss 20b Local & 0.21 & 0.07 & 0.56 & 0.73 \\
\hline
Grok 4.1 fast non-reasoning & 0.35 & 0.21 & 0.56 & 0.62 \\
\hline
Grok 4.1 fast reasoning & 0.32 & 0.13 & 0.59 & 0.7 \\
\hline
Devstral & 0.42 & 0.23 & 0.59 & 0.63 \\
\hline
Magistral Medium & 0.28 & 0.15 & 0.56 & 0.64 \\
\hline
Magistral 24b API & 0.27 & 0.12 & 0.57 & 0.68 \\
\hline
Devstral Small API & 0.31 & 0.2 & 0.55 & 0.6 \\
\hline
Magistral 24b Local & 0.26 & 0.1 & 0.57 & 0.71 \\
\hline
Devstral Small Local & 0.3 & 0.2 & 0.55 & 0.6 \\
\hline
Qwen3 235b thinking & 0.32 & 0.07 & 0.62 & 0.8 \\
\hline
Qwen3 30b thinking API & 0.19 & 0 & 0.59 & 1 \\
\hline
Qwen3 235b instruct & 0.32 & 0.21 & 0.55 & 0.6 \\
\hline
Qwen3 30b instruct API & 0.23 & 0.09 & 0.57 & 0.72 \\
\hline
DeepSeek Chat & 0.36 & 0.27 & 0.54 & 0.57 \\
\hline
DeepSeek Reasoner & 0.35 & 0.15 & 0.59 & 0.68 \\
\hline
DeepSeek R1 14b Local & 0.29 & 0.11 & 0.59 & 0.72 \\
\hline
\end{tabular}%
}
\end{table}

\begin{table}[H]
\centering
\caption{Experiment 5 TP, FP, TN, and FN counts}
\label{tab:exp5-counts}
\resizebox{\columnwidth}{!}{%
\begin{tabular}{|l|c|c|c|c|}
\hline
\textbf{Model} & \textbf{True Positives} & \textbf{False Positives} & \textbf{True Negatives} & \textbf{False Negatives} \\
\hline
GPT Oss 120b & 19 & 9 & 79 & 69 \\
\hline
GPT Oss 20b API & 18 & 7 & 81 & 70 \\
\hline
GPT Oss 20b Local & 19 & 7 & 81 & 69 \\
\hline
Grok 4.1 fast non-reasoning & 31 & 19 & 69 & 57 \\
\hline
Grok 4.1 fast reasoning & 29 & 12 & 76 & 59 \\
\hline
Devstral & 37 & 21 & 67 & 51 \\
\hline
Magistral Medium & 25 & 14 & 74 & 63 \\
\hline
Magistral 24b API & 24 & 11 & 77 & 64 \\
\hline
Devstral Small API & 28 & 18 & 70 & 60 \\
\hline
Magistral 24b Local & 23 & 9 & 79 & 65 \\
\hline
Devstral Small Local & 27 & 18 & 70 & 61 \\
\hline
Qwen3 235b thinking & 29 & 7 & 81 & 59 \\
\hline
Qwen3 30b thinking API & 17 & 0 & 88 & 71 \\
\hline
Qwen3 235b instruct & 29 & 19 & 69 & 59 \\
\hline
Qwen3 30b instruct API & 21 & 8 & 80 & 67 \\
\hline
DeepSeek Chat & 32 & 24 & 64 & 56 \\
\hline
DeepSeek Reasoner & 31 & 14 & 74 & 57 \\
\hline
DeepSeek R1 14b Local & 26 & 10 & 78 & 62 \\
\hline
\end{tabular}%
}
\end{table}

\begin{table}[H]
\centering
\caption{Experiment 5 reliability}
\label{tab:exp5-reliability}
\resizebox{\columnwidth}{!}{%
\begin{tabular}{|l|c|c|c|c|c|}
\hline
\textbf{Model} & \textbf{Timeouts} & \textbf{Invalid Results} & \textbf{Detections} & \textbf{Average runtime (ms)} & \textbf{Errors} \\
\hline
GPT Oss 120b & 0 & 31 & 368 & 25170.26 & 0 \\
\hline
GPT Oss 20b API & 0 & 11 & 181 & 31532.16 & 0 \\
\hline
GPT Oss 20b Local & 17 & 22 & 185 & 91468.28 & 1 \\
\hline
Grok 4.1 fast non-reasoning & 0 & 0 & 539 & 3674.8 & 0 \\
\hline
Grok 4.1 fast reasoning & 0 & 0 & 318 & 23957.17 & 0 \\
\hline
Devstral & 0 & 1 & 633 & 11376.98 & 0 \\
\hline
Magistral Medium & 0 & 16 & 283 & 6339.48 & 0 \\
\hline
Magistral 24b API & 0 & 14 & 274 & 3189.82 & 0 \\
\hline
Devstral Small API & 0 & 4 & 416 & 2031.21 & 0 \\
\hline
Magistral 24b Local & 0 & 12 & 297 & 33693.21 & 0 \\
\hline
Devstral Small Local & 0 & 3 & 436 & 47081.38 & 2 \\
\hline
Qwen3 235b thinking & 0 & 0 & 268 & 101956.21 & 0 \\
\hline
Qwen3 30b thinking API & 0 & 2 & 93 & 24519.21 & 0 \\
\hline
Qwen3 235b instruct & 0 & 1 & 470 & 7489.71 & 0 \\
\hline
Qwen3 30b instruct API & 0 & 9 & 274 & 4488.86 & 0 \\
\hline
DeepSeek Chat & 0 & 0 & 594 & 17888.67 & 0 \\
\hline
DeepSeek Reasoner & 0 & 1 & 363 & 88136.39 & 0 \\
\hline
DeepSeek R1 14b Local & 1 & 3 & 312 & 64570.89 & 0 \\
\hline
\end{tabular}%
}
\end{table}

\end{document}